\documentclass[12pt]{article}
\usepackage{multirow}
\usepackage{graphicx}
\usepackage{amsmath}
\usepackage{amsfonts}
\usepackage{amssymb}
\usepackage{amsthm}

\newtheorem{thm}{Theorem}

\newtheorem{lem}[thm]{Lemma}
\newtheorem{prop}[thm]{Proposition}
\theoremstyle{definition}

\theoremstyle{remark}

\begin{document}

\title{Efficient Estimation of Nonlinear Finite Population \\
Parameters Using Nonparametrics}
\author{{\sc Camelia Goga  }$^1$ \quad and \quad {\sc Anne Ruiz-Gazen}$^2$ \\
$^1$ IMB, Universit\'e de Bourgogne, 
 DIJON - France \\
$^2$ TSE, Universit\'e Toulouse 1,
Toulouse, France.\\
email :  $camelia.goga$@u-bourgogne.fr, $ruiz$@cict.fr }

\maketitle
\date



\baselineskip=16pt 


\begin{abstract} Currently, the high-precision estimation of nonlinear parameters such as Gini indices, low-income proportions or other measures of inequality is particularly crucial. In the present paper, we propose a general class of estimators for such parameters that take into account univariate auxiliary information assumed to be known for every unit in the population. Through a nonparametric model-assisted approach, we construct a unique system of survey weights that can be used to estimate any nonlinear parameter associated with any study variable of the survey, using a plug-in principle. Based on a rigorous functional approach and a linearization principle, the asymptotic variance of the proposed estimators is derived, and variance estimators are shown to be consistent under mild assumptions. The theory is fully detailed for penalized B-spline estimators together with suggestions for practical implementation and guidelines for choosing the smoothing parameters. The validity of the method is demonstrated on data extracted from the French Labor Force Survey.  Point and confidence intervals estimation for the Gini index and the low-income proportion are derived. Theoretical and empirical results highlight our interest in using a nonparametric approach versus a parametric one when estimating nonlinear parameters in the presence of auxiliary information.

\end{abstract}

\vspace{1mm}

\vspace{1mm}

\section{ Introduction} \label{sec:intro}
The estimation of nonlinear parameters in finite populations has become a crucial problem in many recent surveys.
For example, in the European Statistics on Income and Living
Conditions (EU-SILC) survey, several indicators for studying social inequalities and poverty are considered; these include the Gini index, the at-risk-of-poverty rate, the quintile share ratio and the low-income proportion.
Thus, deriving estimators and confidence intervals for such indicators is particularly useful. In the present paper, assuming that we have a single continuous auxiliary variable
available for every unit in the population, we propose a general class of estimators that take into account the auxiliary variable, and we derive their asymptotic properties for general survey designs. The class of estimators we propose is based on a nonparametric model-assisted approach. Interestingly, the estimators can be written as a weighted sum of the sampled observations, allowing a unique weight variable that can be used to estimate any complex parameter associated with any study variable of the survey. Having a unique system of weights is very important in multipurpose surveys such as the EU-SILC survey.

The estimation of nonlinear parameters is a problem that has already been addressed in several papers such as Shao (1994) for L-estimators, Binder and Kovacevic (1995) for the Gini index and Berger and Skinner (2003) for the low-income proportion. We mention also the very recent work of Opsomer and Wang (2011). Taking  auxiliary information into account
for estimating means or totals is a topic that has been extensively studied in the literature; it now encompasses the model-assisted and the calibration approaches, which coincide in particular cases (S\"arndal, 2007). In a model-assisted setting, linear models are usually used, thus leading to the well-known generalized regression estimators (GREG). Some nonparametric models have also been considered (Breidt and Opsomer, 2009). However,
to the best of our knowledge, ratios, distribution functions and quantiles are the only examples of nonlinear parameters estimated using auxiliary information.

To derive our class of estimators and their asymptotic properties, we use an approach based on the influence function developed by Deville (1999). This approach
utilizes a functional interpretation of the parameter of interest and a linearization principle to derive asymptotic approximations of the estimators.
In general, the precision of an estimator $\widehat{\Phi}$ of a nonlinear finite population parameter $\Phi$ is obtained by resampling techniques or linearization approaches and in the present paper we focus on linearization techniques. When a sample $s$ is selected from the finite population $U$ according to a sampling design $p(\cdot)$,
the linearization of $\widehat{\Phi}$ leads under some assumptions, to the following approximation:
\begin{eqnarray}
\hat \Phi-\Phi \simeq \sum_s\frac{u_k}{\pi_k}-\sum_Uu_k\label{lin_Phi}
\end{eqnarray}
where $\pi_k=Pr(k\in s)> 0$ denotes  the first-order inclusion probability for element $k$ under the design $p(\cdot)$. The right term of (\ref{lin_Phi})
is the difference between the well-known Horvitz-Thompson estimator and the parameter it estimates, namely the total of the variable $u_k$ over the population $U$.
Here,  $u_k$ referred to as the linearized variable of $\Phi$ and the way it is derived depends on the type of linearization method used which could include the Taylor series (S\"arndal \textit{et al.}, 1992),  estimating equations (Binder, 1983) or influence function (Deville, 1999) approaches. The artificial variable $u_k$ is used to compute the approximative variance of $\widehat{\Phi}$ as
\begin{eqnarray}
\sum_s\sum_s(\pi_{kl}-\pi_k\pi_l)\frac{u_k}{\pi_k}\frac{u_l}{\pi_l},\label{var_ht_lin}
\end{eqnarray}
with $\pi_{kl}=Pr(k\in s, l\in s)$ the joint inclusion probability for the elements $k,l\in U.$

\begin{figure}[htbp]
\centerline{\includegraphics[width=12cm,height=5.5cm]{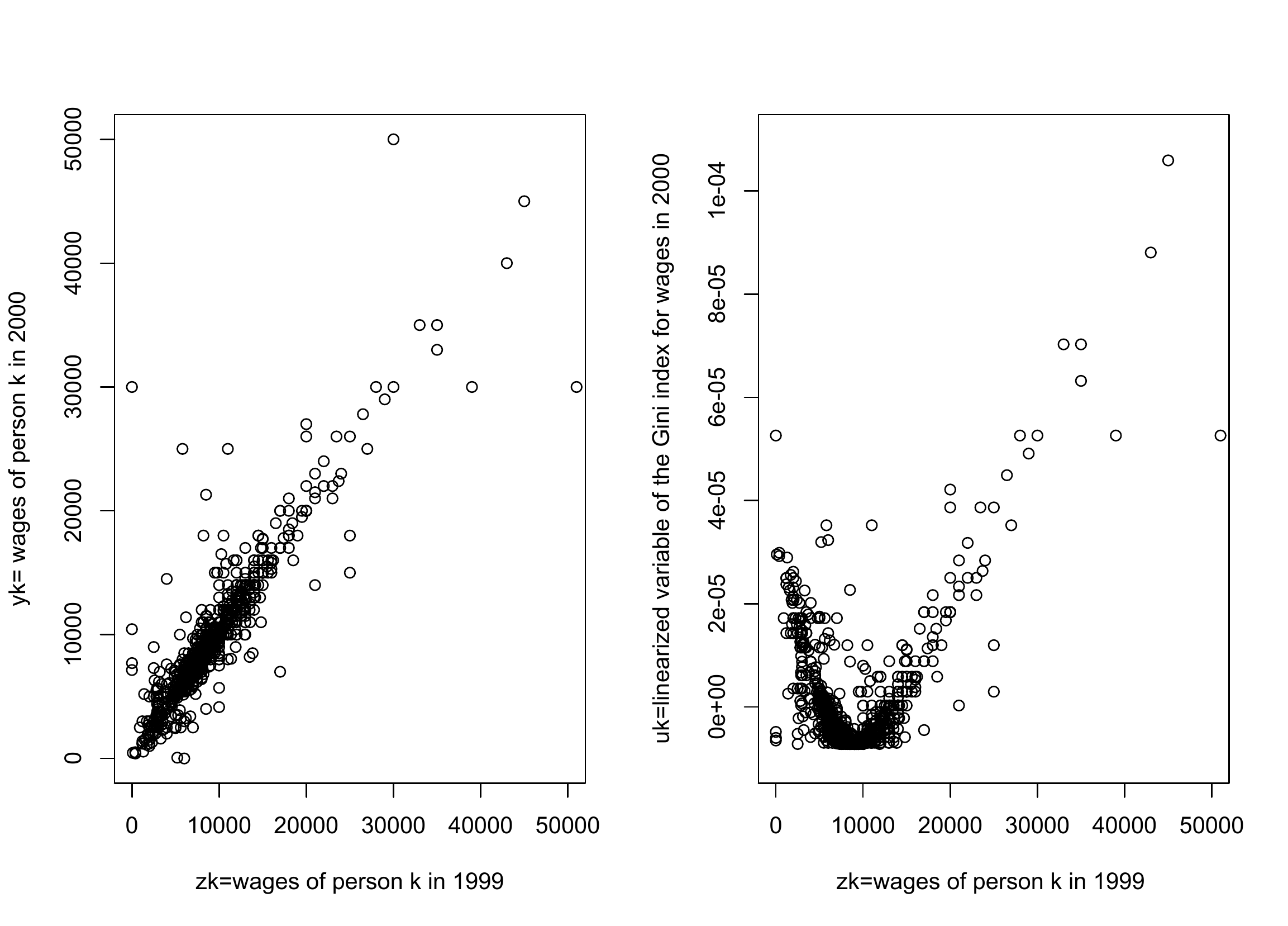}}
\caption{Left plot: $y_k$: the wages of person $k$ in 2000 against $z_k$: the wages of person $k$ in 1999. Right plot:
$u_k$:  linearized variable of the Gini index for the wages in 2000 for person $k$ against $z_k$: the wages of person $k$ in 1999.}
\label{intro1}
\end{figure}

Roughly speaking, when examining (\ref{lin_Phi}) and (\ref{var_ht_lin}), we can see that, if we estimate in an efficient way $\sum_Uu_k$,
we will achieve a small approximative variance and good precision for $\widehat{\Phi}$.
As stated above, it is well known that auxiliary information is useful for improving on the estimation of a total in terms of efficiency and, based on a linear model, the use of a GREG estimator is the most common alternative. When estimating a total, note that the asymptotic variance of the GREG estimator depends on the residuals of the study variable on the auxiliary variable.
Because linearized variables may have complicated mathematical expressions, fitting a linear model onto linearized variables may not be the most appropriate choice. This may occur even if the study and the auxiliary variables have a clear linear relationship, as illustrated in the following example.  Consider a data set of size
1000 extracted from the French Labor Force Survey and consider $y_k$ (the wages of person $k$ in 2000) as the study variable and $x_k$ (the wages of person $k$ in 1999) as the auxiliary variable.
We now consider the problem of estimating the Gini index. The expression of the linearized variable $u_k$, $k\in U$  for the Gini index is given in Binder and Kovacevic (1995) and recalled in equation
(\ref{giniuk}). It is a complex function of the study variable $y_k$, $k\in U$. In the left (resp. right) graphic of Figure \ref{intro1}, the study variable $y_k$ is plotted (resp. the linearized variable $u_k$) on the $y$-axis and the auxiliary variable $x_k$ is plotted on the $x$-axis. The relationship between the study variable and the auxiliary variable is almost linear; however the relationship between the linearized variable of the Gini index and the auxiliary information is no longer linear.
The consequence of this  is that we cannot increase the efficiency of estimating a Gini index if we take  the auxiliary information
into account through a GREG estimator. Therefore, nonparametric models should be preferred to estimate nonlinear parameters $\Phi$.
Recent work already employs nonparametric models to estimate totals (Breidt and Opsomer, 2000, Breidt \textit{et al.}, 2005 and Goga, 2005). The use of nonparametrics prevents model failure; however the improvement over parametric estimation for totals and means may not be significant enough to justify the supplemental difficulties of implementing nonparametric methodology. As illustrated above, the motivation for using nonparametrics becomes much stronger when estimating nonlinear parameters.  Note that the use of nonparametric regression to estimate distribution functions and quantiles has also been  studied, for example in Johnson \textit{et al.} (2008); however, to our knowledge, this has not been performed for other nonlinear parameters.

We propose a novel methodology that allows for the efficient estimation of any parameter $\Phi$ by combining the functional approach (Deville, 1999) with any of the previously suggested nonparametric methods. One issue with the functional approach is that several technical details are not provided in Deville (1999); thus it is difficult to derive rigorous proof of asymptotic results by following this approach.  In the present paper, we propose to clarify some important points and derive rigorous proofs of our asymptotic results. Most importantly, we prove that the total variation distance between finite measures is an adequate choice for the derivation of asymptotic approximations in this context. Asymptotic results are detailed at length for penalized B-spline nonparametric estimators.

The estimators under study combine two types of nonlinearity: nonlinearity due to the expression of a complex parameter and  nonlinearity due to nonparametric estimation. We propose a two-step linearization procedure that provides an approximation of the nonparametric estimator via a Horvitz-Thompson estimator of a total
using an artificial variable. Roughly speaking, this artificial variable corresponds to the residuals of the linearized variable $u_k$ on
the fitted values under the model. Because the linearized variables depend on the parameter of interest, the residuals will also depend on this parameter. The consequence of this important and general property is that the nonparametric approach helps
to get a unique system of weights that may lead to a gain in efficiency for different complex parameters.

The paper is structured as follows: the second section provides some background information on the nonparametric estimation of a finite population total in a general framework. In the third section, a class of nonparametric substitution estimators  based on nonparametric regression is introduced. Variance approximations are derived using the influence function linearization approach (Deville, 1999) in a general nonparametric setting. We propose in the fourth section a  penalized B-spline model-assisted estimator for the finite population totals which is in fact an extension to a survey sampling framework of the penalized B-spline estimator studied in Claeskens \textit{et al.} (2009). We prove that the estimator is asymptotically design-unbiased  and consistent. Next, we build the nonparametric penalized spline estimation for nonlinear parameters
and we assess the validity of the two-step linearization technique. The fifth section defines a  class of consistent variance estimators  while section six contains a case study. The  data set is extracted   from the French Labor Force surveys of 1999 and 2000 as presented previously.  Asymptotic and finite-sample properties of the regression B-spline estimators are illustrated
for the simple random sampling without replacement and the stratified simple random sampling.
This section also includes suggestions for practical implementation and guidelines for choosing the smoothing parameters.
Finally, section seven concludes this study and the assumptions and the technical proofs together with some discussion are provided in the Appendix.

\section{\baselineskip=10pt Nonparametric model-assisted estimation of finite population totals} \label{sec:nonptotal}


We focus on the estimation of the total $t_y=\sum_{k=1}^N y_k=\sum_U y_k$ of the study variable $\mathcal Y$ over $U$,  taking into account the univariate auxiliary variable $\mathcal Z.$ The values $z_1, \ldots, z_N$ of $\mathcal Z$ are assumed to be known for the entire population.

Many approaches can be used to take into account auxiliary information $\mathcal Z$ and thus improve on the Horvitz-Thompson estimator  $\hat t_{y,HT}=\sum_s y_k/\pi_k. $
The goal is to derive a weighted linear estimator $\hat t_{yw}=\sum_sw_{ks}y_k$ of $t_y,$
such that the sample weights $w_{ks}$ do not depend on the study variable values $y_k$ but include the values $z_k,$ for all $k\in U. $  The construction of   the model-assisted (MA) class of estimators $\hat t_{yw}$ is based on a superpopulation model $\xi$:
\vspace{-3mm}
\begin{eqnarray}
\xi : \quad y_k=f(z_k)+\varepsilon_k\label{model}
\end{eqnarray}
where the $\varepsilon_k$ are independent random variables with mean zero and variance $v(z_k). $ If $f(z_k)$ was known for all $k\in U,$ the total $t_y$ may be estimated by the generalized difference estimator (Cassel \textit{et al.}, 1976),
\vspace{-3mm}
 \begin{eqnarray}
\hat t_{y,\mbox{\tiny{diff}}}=\sum_s\frac{y_k-f(z_k)}{\pi_k}+\sum_Uf(z_k).\label{generalizedifer}
\end{eqnarray}
Note that $\hat t_{y,\mbox{\tiny{diff}}}$ consists in the difference between the Horvitz-Thompson estimator $\hat t_{y,HT}$ and its bias under the model $\xi,$ namely $\sum_s f(z_k)/ \pi_k-\sum_Uf(z_k)$. As a consequence, $\hat t_{y,\mbox{\tiny{diff}}}$ is unbiased under the model, $E_{\xi}(\hat t_{y,\mbox{\tiny{diff}}})=t_y$ and moreover, it is unbiased under the sampling design, $E_p(\hat t_{y,\mbox{\tiny{diff}}})=t_y.$ The variance of $\hat t_{y,\mbox{\tiny{diff}}}$ under the sampling design is given by
\begin{eqnarray}
V_p(\hat t_{y,\mbox{\tiny{diff}}})=\sum_U\sum_U(\pi_{kl}-\pi_k\pi_l)\frac{y_k-f(z_k)}{\pi_k}\frac{y_l-f(z_l)}{\pi_l}\label{varHT_diff}
\end{eqnarray}
which shows clearly that  the difference estimator $\hat t_{y,\mbox{\tiny{diff}}}$ is more efficient than the Horvitz-Thompson estimator $\hat t_{y,HT}$ if $f(z_k)$ approximates well $y_k $ for all $k\in U.$

In practice, we don't know the true regression function $f,$ thus we use an estimator of it. Generally, this estimator is obtained using a two-step procedure: we estimate first $f$ by $\tilde f $ under the model $\xi$ and next, we estimate $\tilde f$ by $\hat f$ using  the sampling design. Plugging $\hat f$ in (\ref{generalizedifer}), yields the  final estimator of $t_y.$

 The linear regression function $f(z_k)=\mathbf{z}'_k\mathbf{\beta}$ yields the generalized regression estimator (GREG) extensively studied  by S\"arndal \textit{et al.} (1992). The GREG estimator is efficient if the model fits the data well, but if the model is misspecified, the GREG estimator exhibits no improvement over the Horvitz-Thompson estimator and may even lead to a loss of efficiency.  One way of guarding against model failure is to use nonparametric regression which does not require  a predefined parametric mathematical expression for $f$.

Recently, Breidt and Opsomer (2000) proposed local linear estimators and Breidt \textit{et al.} (2005) and Goga (2005) used nonparametric spline regression. The unknown $f$ function is approximated by the projection of the population vector $\mathbf{y}_U=(y_1, \ldots, y_N)'$ onto different basis functions, such as the basis of truncated $q$th degree polynomials in Breidt \textit{et al.} (2005) and the B-spline basis in Goga (2005).  In the following, we briefly recall the definition and the main asymptotic properties of  nonparametric model-assisted estimators for finite population totals (see also Breidt and Opsomer, 2009).

 Let $\tilde f_{y,k}$ be the estimator of $f(z_k)$ obtained at the population level using one of the three nonparametric methods mentioned above.  Plugging  $\tilde f_{y,k}$ into (\ref{generalizedifer}) results in
 the following nonparametric generalized difference pseudo-estimator of the finite population total:
\begin{eqnarray}
 t^*_{y,\mbox{\tiny diff}} & = &\sum_{ s}\frac{y_k-\tilde f_{y,k}}{\pi_k}+\sum_{ U}\tilde f_{y,k}.\label{estim1}
\end{eqnarray}
 Note that $ t^*_{y,\mbox{\tiny diff}}$ is called a pseudo-estimator because it is not feasible in practice since $\tilde f_{y,k}$ is unknown. This pseudo-estimator is still design-unbiased but it is  model-biased because  nonparametric estimators  $\tilde f_{y,k}$ are  biased for $f(z_k)$ (Sarda and Vieu, 2000). Nevertheless, under supplementary assumptions (Breidt and Opsomer, 2000 and Goga, 2005), the bias under the model vanishes asymptotically to zero when the population and the sample sizes go to infinity. The unknown quantities $\tilde f_{y,k}$ are usually obtained by  least squares methods (ordinary, weighted or penalized) and we may write
\begin{eqnarray}
\tilde f_{y,k}=\mathbf{q}_k'\mathbf{y}_U, \quad \mbox{for all } k\in U\label{estimf}
\end{eqnarray}
where the $N$ dimensional vector $\mathbf{q}_k$ depends on the population values $z_k, $  $k\in U$ as well as on the projection matrix for the considered basis functions, but does not depend on $\mathcal Y.$  The expression of $\mathbf{q}_k$ depends on the  chosen nonparametric method, as discussed in Breidt and Opsomer (2000),  Breidt \textit{et al.} (2005) and Goga (2005). \\
As in the parametric case, we estimate $\tilde f_{y,k}$ by ${\hat f}_{y,k}$ using the sampling design,
\begin{eqnarray}
{\hat f}_{y,k}=\widehat{\mathbf{q}}'_{ks}\mathbf{y}_s, \quad \mbox{for all } k\in U\label{estimhatf}
\end{eqnarray}
where  $\widehat{\mathbf{q}}'_{ks}$ is the $n$-dimensional design-based estimator of $\mathbf{q}'_k$ and $\mathbf{y}_s=(y_k)_{k\in s}$ is the sample restriction of $\mathbf{y}_U.$
 Plugging ${\hat f}_{y,k}$ into (\ref{estim1}) yields the following nonparametric model-assisted estimator (NMA)
 \vspace{-0.3cm}
\begin{eqnarray}
\hat t_{y,np} & = &\sum_{s}\frac{y_k-{\hat f}_{y,k}}{\pi_k}+\sum_{U}{\hat f}_{y,k}.\label{estim2}
\end{eqnarray}
This estimator can be written as a weighted sum  of the sampled observations
\begin{eqnarray}
\hat t_{y,np}=\sum_{s}w_{ks}y_k=\mathbf{w}'_s\mathbf{y}_s, \label{estpondere}
\end{eqnarray}
where the weights $\mathbf{w}_s=(w_{ks})_{k\in s}$ depend only on the sample and on the auxiliary information,
\vspace{-3mm}
\begin{eqnarray}
\mathbf{w}_s  =  \mathbf{\Pi}_s^{-1}\mathbf{1}_s-\widehat{\mathbf{Q}}'_s\mathbf{\Pi}_s^{-1}\mathbf{1}_s+\widehat{\mathbf{Q}}'_U\mathbf{1}_U,\label{expres_gen_w}
\end{eqnarray}
with $\mathbf{1}_s$ the $n$ dimensional vector of ones, $\mathbf{\Pi}_s$  the $n\times n$ diagonal matrix with $\pi_k,$ $k\in s,$ along the diagonal  and  $\widehat{\mathbf{Q}}_U$ the $N\times n$ matrix having $\widehat{\mathbf{q}}'_{ks}$ as rows with sample restriction  $\widehat{\mathbf{Q}}_s=(\widehat{\mathbf{q}}'_{ks})_{k\in s}.$
The estimator (\ref{estpondere}) is a nonlinear function of Horvitz-Thompson estimators, and its asymptotic variance has been obtained on a case-by-case study.
\noindent Under mild hypothesis (Breidt and Opsomer, 2000, Breidt \textit{et al.}, 2005 and Goga, 2005),  $\hat t_{y,np}$ is asymptotically design-unbiased, namely $\mbox{lim}_{N\rightarrow \infty} E_p(\hat t_{y,np}-t_y)/N=0$ and design $\sqrt{n}$-consistent in the sense that
\vspace{-3mm}
\begin{eqnarray}
N^{-1}(\hat t_{y,np}-t_y) & = & O_p(n^{-1/2}).\label{aproximtnp}
\end{eqnarray}
Moreover, it can be approximated by the nonparametric generalized difference estimator $t^*_{y,\mbox{\tiny diff}},$
\vspace{-3mm}
\begin{eqnarray}
N^{-1}(\hat t_{y,np}-t_y) & = & N^{-1}(t^*_{y,\mbox{\tiny diff}}-t_y)+o_p(n^{-1/2}).\label{tclvarlin}
\end{eqnarray}
 Furthermore, if the asymptotic distribution of $V_p(t^*_{y,\mbox{\tiny diff}})^{-1/2}(t^*_{y,\mbox{\tiny diff}}-t_y)$ is normal $\mathcal{N}(0,1)$, we have that  the asymptotic distribution of $V_p(t^*_{y,\mbox{\tiny diff}})^{-1/2}(\hat t_{y,np}-t_y)$ is also normal $\mathcal{N}(0,1)$ where $V_p(t^*_{y,\mbox{\tiny diff}})$ is obtained according to formula (\ref{varHT_diff}) applied to residuals $y_k-\tilde f_{y,k}.$ This means that the NMA estimators bring an improvement over parametric methods and the Horvitz-Thompson estimator when the relation between $\mathcal Y$ and $\mathcal Z$ is not linear. In this case, the residuals $y_k-\tilde f_{y,k}$ will be smaller than under a parametric smoother, which explains the diminution of the design variance of NMA estimators. Nevertheless, nonparametric estimators require that the auxiliary information should be known on the whole population unlike the GREG estimator that requires only the finite population total for $\mathcal Z.$

 The efficiency of NMA estimators depends on the choice of the smoothing parameters. Opsomer and Miller (2005) and Harms and Duchesne (2010) derive the optimal bandwidth for the local polynomial regression, while Breidt \textit{et al.} (2005) circumvent the   issue of the  number of knots by introducing a penalty coefficient. They also give a practical method for estimating this penalty.

\section{\baselineskip=10pt Nonparametric model-assisted estimation\\ of nonlinear finite population parameters} \label{sec:nonpaux}

\subsection{Definition of the nonparametric substitution estimator}
Let us consider the estimation of some nonlinear parameters $\Phi$ by taking into account univariate auxiliary information known for all the population units.
Examples of a nonlinear parameter of interest $\Phi$ include the ratio, the Gini coefficient and the low-income proportion. A parameter $\Phi$ may depend on one or  several variables of interest; however, the same auxiliary variable $\mathcal Z$ will be used to explain these variables of interest. \\
 \noindent We aim to provide a general method for the estimation of  $\Phi$ using $\mathcal Z$ and considering the functional approach introduced by Deville (1999). The methodology consists in
considering a   discrete and finite measure $M=\sum_U\delta_{y_k}$ where $\delta_{y_k}$ is the Dirac measure at the point $y_k$ and $M$ is such that there is  unity mass on each point $y_k$ with $k\in U$ and zero mass elsewhere.  Furthermore, we write  $\Phi$ as a functional $T$ of $M,$
\begin{eqnarray}
\Phi=T(M).\label{phiM}
\end{eqnarray}

\noindent  The nonparametric weights $w_{ks}$ are provided by (\ref{expres_gen_w}) and $M$ is estimated by
$$
\displaystyle \widehat M_{np}=\sum_sw_{ks}\delta_{y_k}.
$$
Even if these weights are derived to estimate the total $t_y, $ they do not depend on the study variable $\mathcal Y$; thus they   can be used to  estimate any nonlinear parameter of interest $\Phi$ when it can be expressed as a function of $M. $ Note that $\widehat M_{np}$ is a random measure of total mass equal to $\hat N_{np}=\sum_sw_{ks}.$\\
 Plugging  $\widehat M_{np}$ into (\ref{phiM}) provides the following nonparametric substitution estimator for $\Phi$,
 $$
 \displaystyle \widehat \Phi_{np}=T(\widehat M_{np}).
 $$
\noindent We will now illustrate the computation of  $\widehat \Phi_{np}$ using the simple case of a ratio $R$  and subsequently
the more intricate case of the Gini index and parameters defined by implicit equations.\\

\noindent\textbf{a.} \textit{The ratio R between two finite population totals}. We write $R=\sum_Uy_k/\sum_Ux_k$ in a functional form as $\displaystyle R=\frac{\int ydM(y)}{\int xdM(x)}. $
The nonparametric estimator of $R$ is easily obtained by replacing the measure $M$ with $\hat M_{np},$ namely $\hat R_{np}=\displaystyle\frac{\int yd\hat{M}_{np}(y)}{\int xd\hat{M}_{np}(x)}=\frac{\sum_sw_{ks}y_k}{\sum_sw_{ks}x_k}.$
A similar estimation of $R$ using GREG weights was previously considered by S\"arndal \textit{et al.} (1992).

\noindent\textbf{b.} \textit{The Gini index}. The Gini index (Nygard and Sandstr\"om, 1985) is given by
$$\mbox{G}=\frac{\sum_Uy_k\left(2F(y_k)-1\right)}{t_y}= \frac{\int (2 F(y)-1) y dM(y)}{ \int y dM(y)}$$
where $F(y)=\int \mathbf{1}_{\{\xi \leq y\}}dM(\xi)/\int dM(y)=\sum_U\textbf{1}_{\{y_k\leq y\}}/N$  is the empirical distribution function. Again, the nonparametric estimator for $G$ is obtained by simply replacing $M$ with $\widehat M_{np}. $ Hence,
\begin{eqnarray}
\widehat{\mbox{G}}_{np} & = & \frac{\sum_sw_{ks}(2\hat F_{np}(y_k)-1)y_k}{\sum_sw_{ks}y_k},\label{giniest}
\end{eqnarray}
where $\hat F_{np}(y)=\displaystyle\frac{\int \mathbf{1}_{\{\xi \leq y\}}d \hat M_{np}(\xi)}{\int d\hat M_{np}(y)}=\frac{\sum_sw_{ks}\mathbf{1}_{\{y_k \leq y\}}}{\sum_sw_{ks}}.$\\

\noindent\textbf{c.} \textit{Parameters defined by an implicit equation}. Let $\Phi$ be defined as the unique solution of an implicit estimating equation  $\sum_{ U} \phi_k(\Phi)=0$ (Binder, 1983) that may be written in a functional form as $\int \phi(\Phi)dM=0. $  We replace $M$ with $\widehat M_{np}$ and the nonparametric sample-based estimator of $\Phi$ is the unique solution of the sample-based estimating equation $\int \phi(\Phi)d\widehat M_{np}=\sum_{s} w_{ks}\phi_k(\widehat \Phi_{np})=0.$ An example of such a parameter is the odds-ratio which is extensively used in epidemiological studies. Goga and Ruiz-Gazen (2012) have studied the estimation of the odds-ratio by taking into account auxiliary information and nonparametric regression.

\subsection{Asymptotic properties of the nonparametric substitution estimator under the sampling design}

In this section, we investigate the asymptotic properties of the nonparametric estimator $\hat \Phi_{np}$, using the asymptotic framework suggested by Isaki and Fuller (1982). Additionally, we make several assumptions (detailed in the Appendix) regarding the regularity of the functional $T$ and the first order inclusion probabilities of the sampling design.

The nonparametric estimator $\widehat\Phi_{np}$ is doubly nonlinear, with nonlinearity due to the parameter $\Phi$ and nonlinearity due to the nonparametric estimation. Our main goal is to approximate $\widehat\Phi_{np}$ using a linear estimator (Horvitz-Thompson type) which will allow to compute the asymptotic variance of $\widehat{\Phi}_{np}. $  This approximation will be accomplished in two steps: first, we will linearize $\Phi$ and next, we will linearize the nonparametric estimator obtained in step one.

The first linearization step is a first-order expansion of $\widehat{\Phi}_{np}$ with the reminder going to zero.  The parameter of interest $\Phi$ is a statistical functional $T$ defined with respect to the measure $M$ or equivalently, with respect to the probability measure $M/N$ (by assumption A1). Using the first-order expansion of statistical functionals $T$ as introduced by  von Mises (1947)  and under the assumption of Fr\'echet differentiability of $T$, the reminder depends on some distance function between $M/N$ and an estimator of this measure (Huber, 1981). Deville (1999) uses these facts to prove the linearization of the Horvitz-Thompson substitution estimator of $\Phi$; however, no details are given about the considered distance, while Goga \textit{et al.} (2009) provide only minimal details. In what follows, we provide a distance between $\widehat M_{np}/N$ and the true $M/N$ which goes to zero when the sample and the population sizes go to infinity.

We consider the total variation distance for two finite and positive measures $M_1$ and $M_2$ to be defined by
$$
d_{\mbox{tv}}(M_1,M_2)=\sup_{h\in \cal{H}} \left|\int h\,dM_1-\int h\,dM_2\right|
$$
with ${\cal H}=\{h:\mathbb{R} \rightarrow \mathbb{R}| \sup_x |h(x)|\leq 1 \}$.
We first prove (lemma \ref{lemth1} from below), that the distance $d_{\mbox{tv}}$ between the Horvitz-Thompson estimator of $M/N$ and the true $M/N$ goes to zero.  Next, we extend the result (lemma \ref{lemth2} from below) to the nonparametric estimator $\widehat{M}_{np}/N.$

\noindent Let $w_{ks}$ represent the Horvitz-Thompson weights, namely $w_{ks}=1/\pi_k$ for all $k\in s$ and let $\widehat M_{HT}=\sum_s\delta_{y_k}/\pi_k$ be the estimator of $M$ using these weights. Let $h\in \mathcal{H}$ and for ease of notation, $x_k=h(y_k)$.
Thus, for all $k\in U,$ $|x_k|\leq 1$ uniformly in $h\in \mathcal{H}$ and
$$
\int h\,d\widehat M_{HT}-\int h\,dM=\sum_s\frac{h(y_k)}{\pi_k}-\sum_Uh(y_k)=\sum_U\left(\frac{I_k}{\pi_k}-1\right)h(y_k),
$$
where $I_k=\mathbf{1}_{\{k\in s\}}$ is the sample membership indicator.
\begin{lem}\label{lemth1}
 Assume (A3) and (A5) from the Appendix. Then,
$$d_{\mbox{tv}}\left(\widehat{M}_{HT}/N,M/N \right)=O_p(n^{-1/2}).$$
\end{lem}
\noindent The proof is provided in the Appendix. We extend now lemma \ref{lemth1} to nonparametric weights  $w_{ks}$ given by (\ref{expres_gen_w}).  Consider again $h\in \mathcal{H}$ and let
\begin{eqnarray*}
\int h\,d\widehat M_{np}-\int h\,d M &= &\sum_sw_{ks}x_k-\sum_Ux_k\\
 & = & \sum_s\frac{x_k-\hat f_{x,k}}{\pi_k}+\sum_U\hat f_{x,k}-\sum_Ux_k
\end{eqnarray*}
where $\hat f_{x,k}$ is obtained from (\ref{estimhatf}) for $y_k$ replaced with $x_k=h(y_k).$ Let also $\tilde f_{x,k}$ obtained from (\ref{estimf}) for $y_k$ replaced with $x_k.$

\begin{lem}\label{lemth2}
 Assume (A3) and (A5) from the Appendix. Assume in addition that: \\
 ($A^*$) for all $k\in U,$ $\frac{1}{N}\sum_U\tilde f^2_{x,k}=O(1)$  uniformly in $h$.\\
 ($A^{**}$) $E_p\left|\frac{1}{N}\sum_U\left(\frac{I_k}{\pi_k}-1\right)(\tilde f_{x,k}-\hat f_{x,k})\right|=O(n^{-1/2})$ uniformly in $h.$ \\
  Then,
$$d_{\mbox{tv}}\left(\widehat{M}_{np}/N,M/N \right)=O_p(n^{-1/2}).$$

\end{lem}
\noindent The proof is provided in the Appendix. In section \ref{sec:bsplines}, we prove that the nonparametric estimator of $M$ constructed using B-spline estimators satisfies the assumptions ($A^*$) and ($A^{**}$) from the above lemma. The results from  Breidt and Opsomer  (2000) may be used to prove the assumptions for local polynomial regression; however, this issue will not be pursued further here.

To provide the first order expansion of $\Phi=T(M),$ we must also define its first derivative. This derivative is referred to as
the influence function  and is  defined as follows (Deville, 1999)
\begin{eqnarray*}
IT(M,y)=\lim_{\varepsilon\rightarrow 0}\frac{T(M+\varepsilon\delta_y)-T(M)}{\varepsilon}
\end{eqnarray*}
where $\delta_y$ is the Dirac measure at point $y$. Note that the above definition is slightly different from the definition of the influence function given by Hampel (1974) in robust statistics, which is based on a probability distribution instead of a finite measure. \\
Let $u_k,$ for all $k\in U$  be   the influence function $IT$ computed at $y=y_k$, namely
$$
u_k=IT(M,y_k), \quad k\in U.
$$
These quantities are referred to as \textit{the linearized variables of $\Phi$} and serve as a tool for computing the approximative variance of $\hat \Phi_{np}. $    They depend on the parameter of interest and they are usually unknown even for the sampled individuals. Deville (1999) provides many practical rules for computing $u_k$ for rather complicated parameters $\Phi.$ \\
\textit{Examples.}  The linearized variable of a ratio $R$ is
\begin{eqnarray}
u_k=\frac{1}{\sum_Ux_k}(y_k-Rx_k)\label{lin_R}
\end{eqnarray}
and for the Gini index,  it is given by
\begin{eqnarray}
u_k=2F(y_k)\frac{y_k-\overline{y}_{k,<}}{t_y}-y_k\frac{1+G}{t_y}+\frac{1-G}{N}\label{lin_G}\label{giniuk}
\end{eqnarray}
where $\overline{y}_{k,<}$ is the mean of $y_j$ lower than $y_k.$\\
  We now provide the main result of this paper. The following theorem is the first linearization step of $\widehat{\Phi}_{np}$. This proves that under broad assumptions  the nonparametric estimator $\widehat\Phi_{np}$ is approximated by the nonparametric estimator for the population total $\sum_Uu_k$ of the linearized variable. The proof is provided in the Appendix.

\begin{thm} (First linearization step) \label{nonlinaux} Assume (A1)-(A3) and (A5) from the Appendix. Additionally assume ($A^*$) and $(A^{**})$ from lemma \ref{lemth2}. Then, the nonparametric substitution estimator $\widehat \Phi_{np}$ fulfills
\begin{eqnarray*}
N^{-\alpha}\left(\widehat\Phi_{np}-\Phi\right) & = &  N^{-\alpha}\left(\sum_sw_{ks}u_k-\sum_Uu_k\right)+o_p(n^{-1/2}).
\end{eqnarray*}
\end{thm}
\noindent We can put $\sum_sw_{ks}u_k$ in the form of an NMA estimator. Let denote $ t^*_{u,np}=\sum_sw_{ks}u_k. $ Using (\ref{expres_gen_w}), we can write
\begin{eqnarray}
t^*_{u,np} & = &\mathbf{w}'_s\mathbf{u}_s=\sum_s\frac{u_k-g^*_{u,k}}{\pi_k}+\sum_Ug^*_{u,k},\label{nonp_u}
\end{eqnarray}
where  $g^*_{u,k}=\widehat{\mathbf{q}}'_{ks}\mathbf{u}_s$ with $\widehat{\mathbf{q}}_{ks}$ is given by (\ref{estimhatf}) and $\mathbf{u}_s=(u_k)_{k\in s}$ is the sample restriction of $\mathbf{u}_U=(u_k)_{k\in U}.$\\
\textbf{Remark 1}: \textit{A model-based interpretation of $ g^*_{u,k}$ may be given. For the nonparametric model $\xi'$,
the linearized variable $u_k$ can be fitted using the auxiliary variable $z_k,$
\begin{eqnarray*}
\xi' : \quad u_k=g(z_k)+\eta_k \label{model2}
\end{eqnarray*}
where the $\eta_k$ are independent random variables with mean zero and variance $\tilde{v}(z_k). $ The estimator of $g$ under the model $\xi',$ denoted by $\tilde g_{u,k}$, is obtained using the same nonparametric method employed for estimating $f$ under the model $\xi.$ This implies  that $\tilde g_{u,k}=\mathbf{q}_k'\mathbf{u}_U$ is the best fit of the population vector $\mathbf{u}_U=(u_k)_{k\in U}$ with $\mathbf{q}_k$ given by  (\ref{estimf}). Furthermore, $\mathbf{q}_k$ is estimated by $\hat{\mathbf{q}}_{ks}$ which leads to the pseudo-estimator   $ g^*_{u,k}=\widehat{\mathbf{q}}'_{ks}\mathbf{u}_s$ of $\tilde g_{u,k}.$  However, unlike the linear case, $ g^*_{u,k}$   is not  an estimate of $\tilde g_{u,k}$ because the  sample linearized variable vector  $\mathbf{u}_s$ is not known and we refer to it as a pseudo-estimator. Remark also that the estimator $\hat\Phi_{np}$ is efficient if the nonparametric model $\xi'$ holds.}

\noindent The nonparametric pseudo-estimator $ t^*_{u,np}$ given by (\ref{nonp_u}) is a nonlinear function of Horvitz-Thompson estimators; however, it estimates a linear parameter of interest, namely the total of $u_k,$ $t_u=\sum_U u_k. $ This indicates that  $ t^*_{u,np}$ is similar to estimators used by   Breidt and Opsomer (2000), Breidt \textit{et al.} (2005) and Goga (2005)  although it is computed for the artificial variable $u_k. $ The second linearization step approximates $ t^*_{u,np}$ by the generalized difference estimator of $\sum_Uu_k$  given by
\begin{eqnarray}
 t^*_{u,\mbox{\tiny diff}} & = & \sum_s\frac{u_k-\tilde g_{u,k}}{\pi_k}+\sum_U\tilde g_{u,k}.\label{diff_u}
\end{eqnarray}

\begin{prop}(Second linearization step)\label{step2}
Assume that $N^{-\alpha} ( t^*_{u,np}- t^*_{u,\mbox{\tiny diff}})= o_p(n^{-1/2}). $ Then,
\begin{eqnarray*}
N^{-\alpha} (t^*_{u,np}-t_u)=  N^{-\alpha} (t^*_{u,\mbox{\tiny diff}}-t_u)+ o_p(n^{-1/2}).
\end{eqnarray*}
\end{prop}
\noindent Based on theorem \ref{nonlinaux} and proposition \ref{step2}, we see that the asymptotic variance of $\widehat\Phi_{np}$ is the variance of $t^*_{u,\mbox{\tiny diff}},$ namely
$$
V_p(t^*_{u,\mbox{\tiny diff}})=\sum_{ U}\sum_{U}\Delta_{kl}\frac{u_k-\tilde g_{u,k}}{\pi_k}\frac{u_l-\tilde g_{u,l}}{\pi_l}.
$$
Moreover, if the asymptotic distribution of $V^{-1/2}_p(t^*_{u,\mbox{\tiny diff}})(t^*_{u,\mbox{\tiny diff}}-t_u)$ is $\mathcal{N}(0,1),$ then the asymptotic distribution of $V^{-1/2}_p(t^*_{u,\mbox{\tiny diff}})(\widehat\Phi_{np}-\Phi)$ is also $\mathcal{N}(0,1).$
\noindent In section \ref{sec:bsplines}, we provide the necessary assumptions for the linearized variables and the auxiliary variable $\mathcal Z$
to obtain an approximation of $ t^*_{u,np}$ by $ t^*_{u,\mbox{\tiny diff}} $ in a B-spline estimation context.\\

\noindent \textbf{Remark 2.} \textit{When the linearized variable $u_k$ is a linear combination of the study variables, the  assumption from proposition \ref{step2} is reduced to assumptions on the study variables.  For example, this occurs in the case of a ratio $R=t_y/t_x,$ where the linearized variable is given by $\displaystyle u_k=\frac{1}{t_x}(y_k-Rx_k)=A_1y_k+A_2x_k.$ The error  $t^*_{u,\mbox{\tiny diff}}-t^*_{u,np}$ can be written as a linear combination of errors between $t^*_{y,\mbox{\tiny diff}}-\hat t_{y,np}$ and  $ t^*_{x,\mbox{\tiny diff}}-\hat t_{x,np}$, respectively.
Using mild regularity assumptions on $\mathcal X,$ $\mathcal Y$ and on the sampling design, $N^{-1}(\hat t_{y,np}- t^*_{y,\mbox{\tiny diff}})$ and $N^{-1}(\hat t_{x,np}-t^*_{x,\mbox{\tiny diff}})$ are shown to be  of order $o_p(n^{-1/2})$ (see Fuller, 2009, for linear regression and section \ref{sec:bsplines} for B-spline estimators). Thus $ t^*_{u,np}-t^*_{u,\mbox{\tiny diff}}$ is also of order $o_p(n^{-1/2})$ provided that $R$ and $N^{-1}t_x$ are bounded.}  \\

\noindent\textbf{Remark 3.} \textit{The asymptotic variance $\widehat\Phi_{np}$ given by theorem \ref{nonlinaux} and proposition \ref{step2} depends on the population residuals $u_k-\tilde g_{u,k}$ of the linearized variables $u_k$ under the model $\xi'$. For the simple case of a  ratio, the relationship between $u_k$ and the study variables is explicit and given by $\displaystyle u_k=A_1y_k+A_2x_k$. If linear models fit the data $x_k$ and $y_k$ well,
then a linear model will also fit $u_k$ well. Nevertheless, for nonlinear parameters such as the Gini index, the relationship between  $u_k$ and the study variable is not as simple as that for the ratio. In such situations,  the use of nonparametric regression methods may provide a major improvement with respect to variance compared to parametric regression.}

\section{\baselineskip=10pt  Penalized B-spline estimators} \label{sec:bsplines}

Spline functions have many attractive properties, and  they are often used in practice due to their good numerical features and ease of implementation. We suppose without loss of generality that all $z_k$ have been normalized and lie in $[0, 1].$ For a fixed $m>1,$ the  set $S_{K,m}$ of spline functions of order $m,$   with  $K$ equidistant interiors knots $0=\xi_0<\xi_1<\ldots<\xi_K<\xi_{K+1}=1$ is the set of   piecewise polynomials of degree $m-1$  that are smoothly connected at the knots (Zhou \textit{et al}., 1998),
$$
S_{K,m}=\{t \in C^{m-2}[0,1]: t(z) \quad \mbox{is a polynomial of degree } (m-1) \mbox{ on each interval} \quad [\xi_i, \xi_{i+1}] \}
$$
For $m=1,$ $S_{K,m}$ is the set of step functions with jumps at knots.
For each fixed set of knots, $S_{K,m}$ is a linear space of functions of dimension  $q=K+m$. A basis for this linear space is provided by the B-spline functions (Schumaker, 1981, Dierckx, 1993)  $B_1, \ldots, B_q$ defined by
\begin{eqnarray}
B_j(x)=(\xi_j- \xi_{j-m})\sum_{l=0}^m\frac{(\xi_{j-l}-x)_+^{m-1}}{\Pi_{r=0, r\neq l}^m(\xi_{j-l}-\xi_{j-r})}\nonumber
\end{eqnarray}
where $(\xi_{j-l}-x)_+^{m-1}=(\xi_{j-l}-x)^{m-1}$ if $\xi_{j-l}\geq x$ and zero, otherwise. For all  $j=1, \ldots, q, $ each function $B_j$ has the knots  $\xi_{j-m}, \ldots, \xi_j$ with $\xi_r=\xi_{\min (\max(r,0), K+1)}$ for $r=j-m,\ldots, j$ (Zhou \textit{et al.}, 1998) which means that its support consists of a small, fixed, finite number of intervals between knots.  Moreover, B-spline are positive functions with a total sum equal to unity:
\begin{eqnarray}
\sum_{j=1}^qB_j(x)=1 \ , \quad \quad x\in [0,1].\label{proprbaza}
\end{eqnarray}
For the same order $m$ and the same knot location, one can use  the truncated power basis  (Ruppert and Carroll, 2000) given by $1, z, z^2, \ldots, z^{m-1}, (z-\xi_1)^{m-1}_+, \ldots, (z-\xi_K)^{m-1}_+$. The B-spline and the truncated power bases are equivalent in the sense that they span the same set of spline functions $S_{K,m}$ (Dierckx, 1993).  Nevertheless, as indicated by Rupert \textit{et al.} (2003), ``the truncated power bases have the practical disadvantage that they are far from orthogonal'', which leads to numerical instability especially if a large number of knots are used.

\subsection{Nonparametric penalized spline estimation for finite population totals}

\vspace{0.3cm}

We now consider the superpopulation model $\xi$ given by (\ref{model}). To estimate the regression function $f,$ we use spline approximation and a penalized least squares criterion.  We define the spline basis vector of dimension $q\times 1$ as  $\mathbf{b'}(z_k)=(B_1(z_k), \ldots, B_q(z_k)), $ $k\in U.$ The penalized spline estimator $\tilde f_{y,k}$ of $f(z_k)$ is given by $
\tilde f_{y,k}=\mathbf{b'}(z_k)\tilde{\boldsymbol{\theta}}_{y,\lambda}$
with $\tilde{\boldsymbol{\theta}}_{y,\lambda}$  as the least squares minimizer  of
\begin{eqnarray}
\sum_{k=1}^N(y_k-\mathbf{b'}(z_k)\boldsymbol{\theta})^{2}+\lambda\int_0^1[(\mathbf{b'}(t)\boldsymbol{\theta})^{(p)}]^2dt,\label{minimizer_theta}
\end{eqnarray}
where $^{(p)}$ represents the $p$-th derivate with $p\leq m-1.$ The solution of (\ref{minimizer_theta}) is a ridge-type estimator,
\begin{eqnarray}
\tilde{\boldsymbol{\theta}}_{y,\lambda}=(\mathbf{B}'_U\mathbf{B}_U+\lambda\mathbf{D}_p)^{-1}\mathbf{B}'_U\mathbf{y}_U,\label{theta_y_lambda}
\end{eqnarray}
where $\mathbf{B}_U$ is the $N\times q$ matrix with rows $\mathbf{b'}(z_k)$ and the $q\times q$  matrix $\mathbf{D}_p$ is  the squared $L^2$ norm applied to the $p$th derivative of $\mathbf{b'}\boldsymbol{\theta}$. Because the derivative of a $B$-spline function of order $m$ may be written as a linear combination of $B$-spline functions of order $m-1$, for equidistant knots  $\mathbf{D}_p=K^{2p}\nabla'_p\mathbf{R}\nabla_p$  (Claeskens \textit{et al.}, 2009) where the matrix $\mathbf{R}$ has elements $R_{ij}=\int_0^1B_i^{(m-p)}(t)B_j^{(m-p)}(t)dt$ with $B_i^{(m-p)}$ as the $B$-spline function of order $m-p$ and $\nabla_p$ as the matrix corresponding to the $p$th order difference operator.

The amount of smoothing is controlled by $\lambda>0.$   The case $\lambda=0$ results in an unpenalized B-spline estimator the asymptotic properties of which have been  extensively studied in the literature (Agarwal and Studden, 1980, Burman, 1991,  and Zhou \textit{et al.}, 1998, among others). The case $\lambda\rightarrow\infty$ is equivalent to fitting a $(p-1)$th degree polynomial.  The theoretical properties of penalized splines with $\lambda>0,$ have been studied only recently by Cardot (2000), Hall and Opsomer (2005), Kauermann \textit{et al.} (2009) and  Claeskens \textit{et al.} (2009).

\noindent The design-based estimators  of $\tilde f_{y,k}$ are
\begin{eqnarray}
\hat f_{y,k}=\mathbf{b'}(z_k) \hat{\boldsymbol{\theta}}_{y,\lambda}\label{estinesantteta}
\end{eqnarray}
where $\hat{\boldsymbol{\theta}}_{y,\lambda}= (\mathbf{B}'_s\mathbf{\Pi}_s^{-1}\mathbf{B}_s+\lambda\mathbf{D}_p)^{-1}\mathbf{B}'_s\mathbf{\Pi}_s^{-1}\mathbf{y}_s$ is the design-based estimator of $\mathbf{\tilde{\boldsymbol{\theta}}}_{y,\lambda}$ and  $\mathbf{B}_s$ is the $n\times q$ matrix given by $\mathbf{B}_s=(\mathbf{b'}(z_k))_{k\in s}.$ We note that $\hat{f}_{y,k}$ may be written as in formula (\ref{estimhatf}) for
$\widehat{\mathbf{q}}'_{ks}=\mathbf{b'}(z_k)(\mathbf{B}'_s\mathbf{\Pi}_s^{-1}\mathbf{B}_s+\lambda\mathbf{D}_p)^{-1}\mathbf{B}'_s\mathbf{\Pi}_s^{-1}.$ \\
Finally, the  $B$-spline NMA estimator of   $t_y$ is as follows:
\begin{eqnarray}
\hat t_{y,BS}& = &\sum_{s}\frac{y_k-\hat f_{y,k}}{\pi_k}+\sum_{U}\hat f_{y,k}\nonumber\\
  & = & \sum_s\frac{y_k}{\pi_k}-\left(\sum_s\frac{\mathbf{b}(z_k)}{\pi_k}-\sum_U\mathbf{b}(z_k)\right)'
  \hat{\boldsymbol{\theta}}_{y,\lambda}.
\label{estimbstot}
\end{eqnarray}
This indicates that $\hat t_{y,BS}$  may be written as a GREG estimator that uses the vectors $\mathbf{b'}(z_k)$ as regressors of dimension $q\times 1$ with $q$ going to infinity and a ridge-type regression coefficient $\hat{\boldsymbol{\theta}}_{y,\lambda}.$ Furthermore, $\hat t_{y,BS}$ is a weighted sum of sampled values $y_k$ with weights $\mathbf{w}_s$ expressed as in (\ref{expres_gen_w}),
\begin{eqnarray}
\mathbf{w}_s=\mathbf{\Pi}_s^{-1}\mathbf{1}_s-\mathbf{\Pi}_s^{-1}\mathbf{B}_s\left(\mathbf{B}'_s\mathbf{\Pi}_s^{-1}\mathbf{B}_s+\lambda\mathbf{D}_p\right)^{-1}\left(\mathbf{B}'_s\mathbf{\Pi}_s^{-1}\mathbf{1}_s-\mathbf{B}'_U\mathbf{1}_U\right). \label{weights_penal_Bsplines}
\end{eqnarray}

\subsubsection*{Regression splines}

\noindent For $\lambda=0,$ we obtained the unpenalized B-spline estimator studied by Goga (2005) and called the regression splines.  The B-spline property given in (\ref{proprbaza})  may be written as $\mathbf{1}'_q\cdot \mathbf{b}(z_k)=1$ with $\mathbf{1}_q$ the $q$ dimensional vector of ones, implying that $\mathbf{1}_s=\mathbf{B}_s\mathbf{1}_q$ and $\mathbf{1}_U=\mathbf{B}_U\mathbf{1}_q.$ Using these two relations in (\ref{weights_penal_Bsplines}) (Goga, 2005), we observe that $\hat t_{y,BS}$ is equal to the finite population total of the prediction $\hat f_{y,k}=\mathbf{b}'(z_k)(\mathbf{B}'_s\mathbf{\Pi}_s^{-1}\mathbf{B}_s)^{-1}\mathbf{B}'_s\mathbf{\Pi}_s^{-1}\mathbf{y}_s,$
$$
\hat t_{y,BS}=\sum_U\hat f_{y,k}=\mathbf{w}'_{s}\mathbf{y}_s
$$ where the weights are given by,
\begin{eqnarray}
\mathbf{w}_{s}=\mathbf{\Pi}^{-1}_s\mathbf{B}_s\left(\mathbf{B}'_s\mathbf{\Pi}_s^{-1}\mathbf{B}_s\right)^{-1}\mathbf{B}'_U\mathbf{1}_U.\label{poidsinfaux}
\end{eqnarray}

Note the similarity with the GREG weights obtained in the case of a linear model when the variance of errors is linearly related to the auxiliary information (S\"arndal, 1980). We note that for a B-spline of order $m=1,$ the estimator $\hat t_{y,BS}$ becomes the well-known poststratified estimator (S\"arndal \textit{et al.}, 1992). \\
Based on assumptions regarding the sampling design and the variable $\mathcal Y,$ (assumptions (A3)-(A5) from the Appendix) and assumptions regarding the distribution of $\mathcal Z$ and the knot number (assumptions (B1)-(B2) in the Appendix), Goga (2005) proved that the B-spline estimator for the total $t_y$ is asymptotically design-unbiased and consistent (equation (\ref{aproximtnp})) and may be approximated by a nonparametric generalized difference estimator (equation (\ref{tclvarlin})). These results are valid  without supplementary assumptions regarding the smoothness of the regression function $f. $

\subsubsection*{Penalized splines using truncated polynomial basis functions}

 Let $\mathbf{c}'(z_k)=\{1, z_k, \ldots, z^{m-1}_k, (z_k-\xi_1)^{m-1}_+, \ldots, (z_k-\xi_K)^{m-1}_+\}$ be the vector basis and let
$\tilde f_{y,k}=\mathbf{c'}(z_k)\mathbf{\tilde\eta}_{y,\rho}$  with $\mathbf{\tilde\eta}_{y,\rho}$  be the least squares minimizer  of $\sum_{k=1}^N(y_k-\mathbf{c'}(z_k)\mathbf{\eta})^{2}+\rho\sum_{j=1}^{K}\eta_{m-1+j}^2$ for $\mathbf{\eta}'=(\eta_0, \ldots, \eta_{m-1+K}). $ The solution is given by
$$
\mathbf{\tilde\eta}_{y,\rho}=(\mathbf{C}'_U\mathbf{C}_U+\rho\mathbf{A})^{-1}\mathbf{C}'_U\mathbf{y}_U
$$
with $\mathbf{C}_U=(\mathbf{c'}(z_k))_{k\in U}$ and the penalty matrix $\mathbf{A}$ having $m-1$ zeros on the diagonal followed by $K$ one values, $\mathbf{A}=\mbox{diag}(0,\ldots,0, 1, \ldots, 1).$ Note that for $\rho=0,$ we obtain the same prediction $\tilde f_{y,k}$ as with an unpenalized B-spline estimation. This results follows from the fact that the two bases are equivalent, thus there exists a square and invertible transition matrix $\mathbf{L}_U$ such that $\mathbf{B}_U=\mathbf{C}_U\mathbf{L}_U$ (Ruppert \textit{et al.}, 2003). For $\rho >0,$ we have $\tilde f_{y,k}=\mathbf{B}'_U(\mathbf{B}'_U\mathbf{B}_U+\rho\mathbf{L}'_U\mathbf{A}\mathbf{L}_U)^{-1}\mathbf{B}'_U\mathbf{y}_U,$ which indicates equivalency to the estimator $\tilde f_{y,k}$ obtained with penalized B-spline fitting given by (\ref{theta_y_lambda}) for $\rho\mathbf{L}'_U\mathbf{A}\mathbf{L}_U=\lambda\mathbf{D}_{q+1}$  (see Claeskens \textit{et al.} (2009) for the expression of $\mathbf{L}_U$ satisfying this equation). \\
In a design-based approach, Claeskens \textit{et al.} (2005)  proved that  the NMA estimator $\hat t_{y,BS}$ is  the population total of the design-based predictions $\hat f_{y,k}=\mathbf{c'}(z_k)(\mathbf{C}'_s\mathbf{\Pi}^{-1}_s\mathbf{C}_s+\rho\mathbf{A})^{-1}\mathbf{C}'_s\mathbf{\Pi}^{-1}_s\mathbf{y}_s. $ They also proved that $\hat t_{y,BS}$ fulfils  properties (\ref{aproximtnp}) and (\ref{tclvarlin}).

\subsection{Asymptotic properties of the B-spline estimator of totals under the sampling design}

In the following, we study the asymptotic properties of $\hat t_{y,BS}$ under the sampling design. We first provide a lemma concerning the convergence of $\hat{\boldsymbol{\theta}}_{y,\lambda}.$ The proofs are based on the results provided by Goga (2005) for the unpenalized B-spline estimator and on the fact that the inverse of the matrix $\frac{1}{N}\mathbf{B}'_U\mathbf{B}_U+\frac{\lambda}{N} \mathbf{D}_p$ is of order $O(K)$ for the penalized B-spline estimator (lemma 1 from Claeskens \textit{et al.}, 2009).

\begin{lem}\label{lemme_thetay}

\begin{itemize}
\item [(a)] Assume assumptions (A4)-(b) and  (B1), (B2)-(a) and (B3) from the Appendix. Then, $||\tilde{\boldsymbol{\theta}}_{y,\lambda}||=O(K^{1/2}).$
\item [(b)] Assume assumptions (A3), (A4)-(b), (A5) and  (B1)-(B3) from the Appendix. Then,
$$
E_p(||\hat{\boldsymbol{\theta}}_{y,\lambda}-\tilde{\boldsymbol{\theta}}_{y,\lambda}||^2)=O\left(\frac{K^3}{n}\right).
$$
\end{itemize}
where $||\cdot||$ is the usual euclidian norm.
\end{lem}
\noindent The proof is provided in the Appendix. We note that for B-spline functions of order $m=1$ and $\lambda=0,$ we obtain a  poststratified estimator with a number of poststrata going to infinity.
In this context, lemma \ref{lemme_thetay}, (b) provides a detailed theoretical justification for the poststratification example in Deville (1999, p. 196).
We note also that to obtain the convergence of $\hat t_{y,BS},$ Claeskens \textit{et al.} (2005) assume that the result from lemma \ref{lemme_thetay}, (b) holds. Finally, we note that GREG estimators may be viewed as a special case when the number of knots is fixed. Papers dealing with this issue usually assume that the regression coefficient satisfies the results from the above lemma (see for example Robinson and S\"arndal, 1983, or Isaki and Fuller, 1983).  A similar result was proved by Cardot \textit{et al.} (2012).

Using lemma \ref{lemme_thetay}, we derive the following results.

\begin{prop}\label{lemma_hattBS} Assume assumptions (A3), (A4)-(b), (A5) and (B1)-(B3) from the Appendix. Then,
\begin{itemize}
\item [(a)] $E_p\left|\frac{1}{N}\left(\hat t_{y,BS}-t_y\right)\right|=O((K/n)^{1/2})). $
\item [(b)] $\frac{1}{N}\left(\hat t_{y,BS}-t_y\right)=\displaystyle{\frac{1}{N}\left( t^*_{y,diff}-t_y\right)+o_p(n^{-1/2})}$
where $t^*_{y,diff}=\sum_{s}\frac{y_k-\tilde f_{y,k}}{\pi_k}+\sum_{U}\tilde f_{y,k}.$
\end{itemize}
\end{prop}
\noindent The proof is provided in the Appendix. Using the Markov inequality,  we see from the first point of proposition \ref{lemma_hattBS} that $\hat t_{y,BS}$ is asymptotically design-unbiased for $t_y$  and $\sqrt n$-consistent as $(\hat t_{y,BS}-t_y)/N=O_p(n^{-1/2}).$ The second point provides an approximation of
$\hat t_{y,BS}$ by the nonparametric generalized difference estimator $ t^*_{y,diff}. $

\subsection{ Calibration with penalized splines}
\vspace{0.3cm}
The spline approach has some interesting  calibration properties. Under the  unpenalized B-spline framework,
  the weights $w_{ks}$ given by (\ref{poidsinfaux}) satisfy the calibration equation for the known population total of B-spline functions, namely
 $$
 \sum_{s}w_{ks}B_j(z_k)=\sum_U B_j(z_k), \quad \mbox{for all } j=1, \ldots, q.
 $$
This relation is easily  obtained using (\ref{proprbaza}) (Goga, 2005). Because the  spline space $S_{K,m}$ is spanned by the B-spline functions $B_j$,  these weights will be calibrated to the total of any polynomial  $z^r$ of degree $r\leq q=K+m. $ In particular, $\sum_s w_{ks}=N$ and $\sum_s w_{ks}z_k=\sum_U z_k.$ Claeskens \textit{et al.} (2005) prove that using the penalized splines and the truncated polynomial basis functions l provides weights that are also calibrated for the finite population totals of the polynomial basis functions $1, z, z^2, \ldots, z^{m-1}. $

\subsection{Nonparametric penalized spline estimation for nonlinear parameters}

\vspace{0.3cm}

We now consider the nonlinear parameter $\Phi$ estimated by $\hat \Phi_{BS}=T(\widehat M_{BS})$
with $\widehat M_{BS}=\sum_sw_{ks}\delta_{y_k}$  and the weights $w_{ks}$ given by (\ref{weights_penal_Bsplines}). As in section \ref{sec:nonpaux}, to linearize $\hat \Phi_{BS}$ we use a two-step procedure. The first-step linearization is given in  theorem \ref{nonlinaux} provided that the assumptions $(A^*)$ and $(A^{**})$ from lemma (\ref{lemth2}) are fulfilled.  These assumptions are crucial because they ensure the convergence of some nonparametric estimator of $M$ to the true measure $M$ according to the distance $d_{tv}.$ Using classical assumptions from a B-spline framework (assumptions (B1)-(B3) from the Appendix) and mild assumptions regarding the sampling design (assumptions (A3) and (A5) from the Appendix), we prove in theorem \ref{thm:bspline} below that  $(A^*)$ and $(A^{**})$ are verified.  The proof is basically based on lemma \ref{lemme_thetay} and the fact that the distance $d_{tv}$ is defined for uniformly bounded functions $h\in \mathcal H,$ ensuring  that the assumption (A4)-(b) is automatically fulfilled.

By conducting this first linearization step, we see that the nonparametric B-spline estimator $\hat \Phi_{BS}$ will be approximated by  the nonparametric B-spline estimator of the total of the linearized variables $u_k$ given by
$$
 t^*_{u,BS}=\mathbf{w}'_s\mathbf{u}_s=\sum_{s}\frac{u_k- g^*_{u,k}}{\pi_k}+\sum_{U}g^*_{u,k},
$$
where $g^*_{u,k}=\mathbf{b'}(z_k)\hat{\boldsymbol{\theta}}_{u,\lambda}\label{estimgvarlin}$
with  $\hat{\boldsymbol{\theta}}_{u,\lambda}=(\mathbf{B}'_s\mathbf{\Pi}_s^{-1}\mathbf{B}_s+\lambda\mathbf{D}_p)^{-1}\mathbf{B}'_s\mathbf{\Pi}_s^{-1}\mathbf{u_s}.$

The second-step linearization  consists of providing an approximation of $ t^*_{u,BS}$ by a nonparametric generalized difference estimator,
$$
 t^*_{u,\mbox{\tiny diff}} = \sum_{s}\frac{u_k-\tilde g_{u,k}}{\pi_k}+\sum_{U}\tilde g_{u,k}.
$$
where  $\tilde g_{u,k}=\mathbf{b'}(z_k)(\mathbf{B}'_U\mathbf{B}_U+\lambda\mathbf{D}_p)^{-1}\mathbf{B}'_U\mathbf{u_U}.$ To obtain this result, we state in theorem \ref{thm:bspline}, (b) a supplementary assumption regarding the linearized variable $u_k.$ Goga and Ruiz-Gazen (2012) prove that the linearized variable $u_k$  of the odds-ratio  satisfies this assumption.

\begin{thm}\label{thm:bspline} Suppose that the sampling design satisfies assumptions (A3) and (A5).   In addition,  assume that (B1)-(B3) hold.
\begin{itemize}
\item [(a)] Assumptions $(A^*)$ and $(A^{**})$ from lemma \ref{lemth2} are fulfilled. \\
As a consequence,  $d_{tv}(\widehat M_{BS}/N,M/N)=O_p(n^{-1/2}).$   Moreover, if the functional $T$ satisfies (A1) and (A2), then
$N^{-\alpha}(\hat \Phi_{BS}-\Phi)=N^{-\alpha}(\hat t_{u,BS}-t_u)+o_p(n^{-1/2}).$
\item [(b)] Suppose that the linearized variables are such that  for all $k\in U,$  $N^{-\alpha+1}u_k$ satisfy (A4)-(b).  Then, $
N^{-\alpha}\left(\hat t_{u,BS}-t_u\right)=N^{-\alpha}\left(\hat t_{u,\mbox{\tiny diff}}-t_u\right)+o_p(n^{-1/2}).\label{approxbspline}$
\end{itemize}
\end{thm}
The proof is provided in the Appendix.

\section{\baselineskip=10pt Variance estimation} \label{sec:multaux}

In this section we undertake a detailed study of the variance estimation  of $\widehat \Phi_{np}. $  We first give  the functional form of the variance of  $\hat t_{y,HT}$ as well as  of its variance estimator  and we propose a variance estimator for $\widehat \Phi_{np}$ and assumptions under which this estimator is consistent.

The Horvitz-Thompson variance $V_{p}(\hat t_{y,HT})=\sum_U\sum_U\Delta_{kl}(y_k/\pi_k)(y_l/\pi_l)$ for $\Delta_{kl}=\pi_{kl}-\pi_k\pi_l$ is a quadratic form that can be written as a functional of some finite and discrete measure. We can write  the variance as follows (Liu and Thompson, 1983),
\begin{eqnarray}
V_{p}(\hat t_{y,HT})=\sum_{(k,l)\in U^*} \psi(y_k,y_l)\label{batchvariance}
\end{eqnarray}
where $U^*=\{(k,l), k, l=1, \ldots N\}$ and $\psi(y_k,y_l)=\Delta_{kl}(y_k/\pi_k)(y_l/\pi_l)$ is a bilinear function of $y_k$ and $y_l.$ It follows from (\ref{batchvariance}), that the Horvitz-Thompson variance $V_{p}$ is the finite population total of $\psi(y_k,y_l)$ over the derived synthetic population $U^*$ of size $N^*=N^2.$ This variance can be put in a functional form as follows
$$
V_{p}(\hat t_{y,HT})=T^*(M^*)=\int \psi(y,y)d M^*(y,y)
$$
where $M^*=\sum_{(k,l)\in U^*}\delta_{(y_k,y_l)}. $ Note that $T^*$ is a functional of degree $1$ with respect to  $M^*,$ namely $T^*(M^*/N^*)=T^*(M^*)/N^*.$
A sample  in this population $U^*$ is $s^*=\{(k,l), k, l\in s\}$ and has size $n^*=n^2. $ Moreover,  the first-order inclusion probabilities over the synthetic population $U^*$ are $\pi^*_{(k,l)}=\pi_{kl}$, which are  exactly the second-order inclusion probabilities with respect to the initial sampling design $p(s).$ The measure $M^*$ is estimated on $s^*$ by $\widehat M^*=\sum_{(k,l)\in s^*}\delta_{(y_k,y_l)}/\pi_{kl}=\sum_{s^*}w^*_{(kl)}\delta_{(y_k,y_l)}$ where $w^*_{(kl)}=1/\pi_{kl}. $ The resulting estimator of  $V_{p}(\hat t_{y,HT})$
 is as follows
 $$
 \widehat V_{p}(\hat t_{y,HT})=T^*(\widehat M^*)=\int \psi(y,y)d \widehat M^*(y,y)=\sum_{(k,l)\in s^*}\frac{\Delta_{kl}}{\pi_{kl}}\frac{y_k}{\pi_k}\frac{y_l}{\pi_l}.
 $$
This is exactly the Horvitz-Thompson variance estimator, as $\sum_{(k,l)\in s^*}$ is equal to $\sum_{k\in s}\sum_{l\in s}. $ Moreover, the functional $T^*$ is Fr\'echet differentiable, with first derivative given by $IT^*(M^*,y)=\psi(y,y). $\\

\noindent  Consider now the asymptotic  variance $AV_p(\widehat{\Phi}_{np})$ of $\widehat{\Phi}_{np}$ given by
\begin{eqnarray}
AV_p(\widehat{\Phi}_{np})=\displaystyle\sum_{U}\sum_{U}\Delta_{kl}\frac{u_k-\tilde g_{u,k}}{\pi_k}\frac{u_l-\tilde g_{u,l}}{\pi_l}\label{varnonlin}
\end{eqnarray}
where $u_k$ is the linearized variable of $\Phi$ and $\tilde g_{u,k}=\mathbf{q}'_k\mathbf{u}_U$ for $\mathbf{u}_U=(u_k)_{k\in U}. $ We recognize  the Horvitz-Thompson variance of the total of the population residuals $e_{ks}=u_k-\tilde g_{u,k}. $ We suggest estimating the variance of $\widehat{\Phi}_{np}$ by using the Horvitz-Thompson variance estimator  with $u_k$ replaced by the sample estimators $\hat u_k,$
\begin{eqnarray}
\widehat {V}_p(\widehat{\Phi}_{np})=\displaystyle\sum_{ s}\sum_{s}\frac{\Delta_{kl}}{\pi_{kl}}\frac{\hat {u}_k-\hat g_{\hat u,k}}{\pi_k}\frac{\hat u_l-\hat g_{\hat u,l}}{\pi_l}\label{varestimnonlin}
\end{eqnarray}
where
$\hat g_{\hat u,k}=\hat{\mathbf{q}}'_{ks}\mathbf{\hat u}_s$ is the sample estimate of $\tilde g_{u,k}=\mathbf{q}'_{k}\mathbf{u}_U. $ The Horvitz-Thompson variance estimator with true linearized variables given by
\begin{eqnarray}
\widehat{AV}_{p}(\widehat{\Phi}_{np})=\sum_{ s}\sum_{ s}\frac{\Delta_{kl}}{\pi_{kl}}\frac{u_k-\tilde g_{u,k}}{\pi_k}\frac{u_l-\tilde g_{u,l}}{\pi_l}. \label{varht}
\end{eqnarray}
The three expressions of variance above depend on the population fits residuals $e_{ks},$ for all $k\in U.$ It follows that we may write $AV_p(\widehat{\Phi}_{np})$ as a functional of $M^*$ depending on parameter $\mathbf{e}_U=(e_{ks})_{k\in U},$

 $$
 AV_p(\widehat{\Phi}_{np})=T^*(M^*,\mathbf{e}_U).
 $$
Furthermore, the Horvitz-Thompson estimator $\widehat{AV}_{p}(\widehat{\Phi}_{np})$ and the variance estimator $ \widehat {V}_p(\widehat{\Phi}_{np})$ can be treated in a functional form as follows
 $$
 \widehat{AV}_{p}(\widehat{\Phi}_{np})=T^*(\widehat  M^*,\mathbf{e}_U), \quad \widehat {V}_p(\widehat{\Phi}_{np})=T^*(\widehat M^*,\mathbf{\hat e}_U).
 $$
Note that   $\mathbf{\hat e}_U=(\hat e_{ks})_{k\in U}$ is the vector of sample-based fit residuals with $\hat e_{ks}=\hat {u}_k-\hat g_{\hat u,k}, $ for all $k\in U. $ Theorem 3 from Goga \textit{et al.} (2009) allows us to establish under additional assumptions that the variance estimator (\ref{varestimnonlin}) is $n$-consistent for the asymptotic variance.

\begin{thm}
Assume that assumptions (A3) and (A5) from the Appendix hold. Also assume  that $N^{1-\alpha}e_{ks}=O(1)$ holds uniformly in k  and
$\displaystyle{nN^{-2\alpha}\sum_U(\hat e_{ks}-e_{ks})^2=o_p(1)}.$  If the Horvitz-Thompson variance estimator $\widehat{AV}_{p}(\widehat{\Phi}_{np})$ is $n$-consistent for $AV_{p}(\widehat{\Phi}_{np}),$ then the variance estimator $ \widehat {V}_p(\widehat{\Phi}_{np})$ is also $n$-consistent for $AV_p(\widehat{\Phi}_{np})$ in the sense that $
nN^{-2\alpha}(\widehat {V}_p(\widehat{\Phi}_{np})-AV_p(\widehat{\Phi}_{np}))=o_p(1).$
\label{estim_var_nonparam}
\end{thm}
\noindent The proof is given in the Appendix. Note that because the functional $T^*$ is Fr\'echet differentiable, the $n$-consistency of the Horvitz-Thompson estimator  $\widehat{AV}_{p}(\widehat{\Phi}_{np})$ for $AV_p(\widehat{\Phi}_{np})$ may also be derived with assumptions on fourth moment of $e_{ks}$ and on fourth-order inclusion probabilities. The reader is referred to Breidt and Opsomer (2000) for additional details.
\section{\baselineskip=10pt Empirical results}\label{section:empirical results}

Let us consider a data set from the French Labor Force
surveys of 1999 and 2000 as the finite populations of interest.
The data consist of the monthly wages (in euros) of 19,378
wage-earners who were sampled in both years. The study variable $y_k$ (resp. the auxiliary variable $x_k$) is the wage of person $k$ in 2000
(resp. 1999). The objective of the simulation studies is to investigate the finite-sample performance of the regression spline estimators for two nonlinear parameters of interest and two different survey designs.
We concentrate in practice on the simple approach of regression B-splines and do not consider the penalized B-splines with $\lambda>0$. The empirical study of penalized splines raises the problem of estimating the parameter $\lambda$ which is beyond the scope of the present paper.
We illustrate the efficiency of the regression B-splines estimators compared to other estimators,
and we also confirm the possibility of conducting valid inference using variance estimators as detailed in the previous section.

The parameters to estimate include the mean, the Gini index and the poverty rate for the wages in 2000 using the wages in 1999 as auxiliary information. The poverty rate is the proportion of individuals whose wages are below the threshold of 60\% of the median wage and correspond to the low-income proportion studied in Berger and Skinner (2003). The Gini index and the low-income proportion are the complex parameters to be estimated and we provide results for the mean as a benchmark. Note that details on the low-income proportion estimator and its associated linearized variable can be found
in Berger and Skinner (2003) and are not provided in the present paper.
In subsection \ref{sec:srswor}, we focus on simple random sampling without replacement and in subsection \ref{sec:stsrswor}, we focus on a stratified simple random sampling without replacement. We consider the following estimators for each parameter:\\
- the Horvitz-Thompson estimator (HT), which does not incorporate any auxiliary information,\\
- poststratified estimators (POST) with a different number of strata bounded at the empirical quantiles for 1999 wages,\\
- the GREG estimator (GREG), which takes into account the 1999 wages as auxiliary information using a simple linear model,\\
- B-spline estimators (BS($m$) where $m$ denotes the spline order), which take into account the wages from 1999 as auxiliary information by using a nonparametric model with different numbers of knots ($K$) located at the quantiles of the empirical distribution for wages from 1999. The $m=2$ and $m=3$  orders are considered.

The poststratified estimator is an example of a B-spline estimator with order $m=1$. The number of strata correspond
to the number of interior knots $K$ plus one.

To use the regression B-spline estimators we propose in a complex survey, and derive confidence intervals, the user must be able to calculate the weights given in equation  (\ref{poidsinfaux}) and the residuals $\hat {u}_k-\hat g_{\hat u,k}$ of equation (\ref{varestimnonlin}).
The weights depend on a spline basis that is easy to obtain using for instance the \texttt{transreg} procedure in the SAS software or the functions \texttt{spline.des} or \texttt{bs} from the \texttt{splines} package  in the R software. Then, it is possible to use standard calibration algorithms by simply providing the ($m+K$) B-spline basis functions as auxiliary variables for calculating the calibrated weights that correspond to equation (\ref{poidsinfaux}). These weights are needed to calculate the substitution estimator of the parameter of interest (e.g. the expression (\ref{giniest}) for the Gini index). To estimate the variance, the linearized variables associated with the parameter have to be estimated. For several inequality indicators, including the Gini index and the low-income proportion, some SAS macro programs are freely available on the web site of Xavier d'Haultf\oe uille. Similar functions are available in the R language upon request from the authors of the present paper. Once the linearized variable is estimated, the residuals of this variable against the auxiliary variable using regression splines are calculated; this can be accomplished  with the \texttt{transreg} procedure in the SAS software. Then, by using the residuals as if they were the study variable in standard variance estimation tools for complex surveys, the user can obtain the estimated approximative variance and derive confidence intervals.

For each simulation scheme, we draw $NS$ samples according to the sampling design and compare the finite-sample properties of the HT estimator, the GREG estimator, the POST and the BS(2) and BS(3) estimators. We set  knots at the quantiles of the empirical distribution of the auxiliary variable in the sample. We also compared the results with  knots set at the quantiles of the empirical distribution of the auxiliary variable over the entire population. Both results are very similar; thus,  we report only on the first method. For the POST, BS(2) and BS(3) estimators we tried different numbers of knots $K$ but only report the results for $K=2$ and $K=4$. Note that in the tables, the results for $K=2$ and $K=4$ are reported in the same columns and separated by a dash.
 For the poststratified estimator, $K=2$ (resp. $K=4$) corresponds to 3 (resp. 5) strata. To summarize, in the following, we compare eight estimators (HT, GREG and  POST, BS(2) and BS(3) with $K=2$ and $K=4$).

There are several ways to estimate the linearized variable (see section \ref{sec:multaux}). In this section, the results are almost the same, regardless of whether we use the simple HT weights, the GREG weights or the B-spline weights for estimating the linearized variable.
We recommend using the simplest weights (that is, the HT weights), which is what we do in the present study.

Estimators  performance of $\hat\theta$
for a parameter $\theta$ is evaluated using the following Monte-Carlo measures:
\begin{itemize}
\item Relative bias in percentage: $\displaystyle \mbox{RB}=\frac{100}{NS} \times \sum_{i=1}^{NS}(\hat\theta_i-\theta) / \theta$.
\item Ratio of root mean squared errors in percentage:
$$\mbox{RRMSE}=100\times \sqrt{\sum_{i=1}^{NS}(\hat\theta_i-\theta)^2}
/ \sqrt{\sum_{i=1}^{NS}(\hat\theta_{i,HT}-\theta)^2}.$$
\item Monte-Carlo Coverage probabilities for a nominal coverage probability of 95\%.
\end{itemize}

\subsection{Simple random sampling without replacement}\label{sec:srswor}

The first survey design we consider is simple random sampling without replacement with three sample sizes ($n=200$, $n=500$ and $n=1000$).
The number of simulations is $NS=$3,000.
The eight estimators are compared and relative biases  and ratios of the roots of the mean squared errors
are provided in Table \ref{compsi} for the different parameters and sample sizes.

Not surprisingly, for complex parameters, the largest efficiency gain is observed  when the B-spline estimators are compared to the HT estimator without auxiliary information. Because the wages from 2000 are almost linearly related to the wages from 1999, considering the B-spline estimator instead of the GREG estimator does not improve the performance of the mean estimation. However, regarding the Gini index and the low-income proportion, the incorporation of auxiliary information using GREG estimators does not improve efficiency compared to the HT estimator while using a B-spline approach improves the results especially for spline functions of order $m=2$.
When comparing the POST estimator with the BS(2) and BS(3) estimators we notice that there is quite a large gain in efficiency when order $m=2$ is used instead of $m=1$, while there is an efficiency loss when $m=3$ is used instead of $m=2$, especially for sample sizes smaller than 1,000.
Moreover, for $m=2$ and $m=3$, the results do not depend heavily on the number of knots and are similar for $K$ between 2 and 4 while for the poststratified estimator, there are large variations in the results, regardless of whether we consider 3 or 5 strata.
The coverage probabilities in table \ref{probsi} illustrate that valid inference can be carried out using B-spline estimators as long as the spline order is not too high,
especially when the sample size is not very large. No problems are detected for B-splines of order $m=1$ and order $m=2$ even when the sample size is $n=200$; however for $m=3$
and $n=200$, the coverage probabilities for the Gini
index estimation are approximately 75\% which is quite far from the 95\% nominal probability. This result indicates that for a moderate sample size, the variance may be underestimated when the order of the splines is larger than two. The results are not given for $m=4$ but we have observed that the problem worsens when we increase the order of the splines. This is not really surprising due to double linearization and nonparametric estimation.
\begin{center}
\begin{table}
\caption{\label{compsi} RRMSE (RB) of HT, GREG and POST, BS(2)
and BS(3) with  $K=2$ - 4 for the mean, the Gini index and the low-income proportion}
\centering
\fbox{
\begin{tabular}{*{7}{c}}
\em Parameter & $n$&\em  HT&\em GREG& \em POST & \em BS(2)  & \em BS(3) \\
\hline
Mean & 200 &  100 (0) & 38 (0) &  71 (0)  - 63 (0)  &  38 (0) - 37 (0)  & 39 (0)  - 41 (0)\\
& 500 &  100 (0) & 40 (0) &  73 (0)  - 65 (0)  &  40 (0) - 39 (0)  & 38 (0)  - 39 (0)\\
 & 1,000 & 100 (0) & 40 (0) &  73 (0)  - 66 (0) & 40 (0) - 40 (0) & 38 (0) - 39 (0) \\
\hline
Gini index & 200 &  100 (1) & 96 (1) & 92 (1)   - 80 (1) & 53 (2) - 53 (2) & 70 (3) - 70 (3) \\
& 500 &  100 (1) & 93 (0) & 93 (1)   - 85 (1) & 50 (1) - 50 (1) & 59 (1) - 56 (1) \\
 & 1,000 & 100 (0) & 92 (0) & 93 (0) - 86 (0) & 49 (0)  - 48 (0)& 55 (1) - 51 (1) \\
\hline
Poverty rate & 200 &  100 (2) & 95 (0) & 92 (0) - 80 (0) & 65 (1) - 65 (1) & 72 (1) - 63 (1)  \\
& 500 &  100 (0) & 95 (0) & 88 (0) - 78 (0) & 64 (0) - 64 (0) & 68 (0) - 62 (0)  \\
& 1,000 & 100 (1) & 94 (0) & 89 (0) - 78 (0) & 64 (0) - 64 (0)& 67 (0) - 61 (0) \\
\hline
\end{tabular}
}
\end{table}
\end{center}
\begin{center}
\begin{table}
\caption{\label{probsi} Coverage probabilities (in \%) for HT, GREG and POST, BS(2)
and BS(3) with  $K=2$ - 4}
\centering
\fbox{
\begin{tabular}{*{7}{c}}
\em Parameter & $n$&\em  HT&\em GREG& \em POST & \em BSPL(2)  & \em BSPL(3)\\
\hline
Mean & 200 &  94 & 95 & 93 - 92& 93 - 93 & 90 - 88 \\
& 500 &  95 & 94 & 93 - 94& 93 - 93 & 91 - 91 \\
 & 1,000 & 95 & 95 & 94 - 93 & 94 - 94 & 93 - 93 \\
\hline
Gini index & 200 & 94 & 93 & 94  - 94 & 89 - 87& 74 - 75 \\
& 500 & 93 & 93 & 93  - 94 & 91 - 90& 83 - 85 \\
 & 1,000 & 95 & 94 & 95 - 94& 94 - 93 & 88 - 90 \\
\hline
Poverty rate & 200 & 94 & 95 & 95  - 95& 95 - 94 &  94  - 94 \\
& 500 & 93 & 95 & 95  - 94& 95 - 95 &  96  - 95 \\
 & 1,000 & 94 & 95 & 96 - 96 & 95 - 96 & 96 - 95\\
\hline
\end{tabular}}
\end{table}
\end{center}
\vspace{-25mm}
\subsection{Stratified simple random sampling without replacement}\label{sec:stsrswor}

For each simulation, we draw $NS=$5,000 samples from the French Labor Force population according to a stratified simple random sampling design without replacement.
We compare the finite-sample properties of the eight estimators considered in the previous subsection.
The strata are spatial divisions of the French territory in six ``regions'' that correspond to the major socio-economic regions of metropolitan France as defined by Eurostat. These regions are the first level of the nomenclature of territorial units for statistics classification (NUTS-1). For our example, we grouped the Northern and Eastern regions together and we grouped the Mediterranean and the Southwestern regions together. The sample size inside each stratum is 200 making the total sample size 1200.
Thus, we used an unequal probability design with a sample rate inside the strata that varied from 5 to 9.3\%.

As previously described, we set the knots at the quantiles of the empirical distribution of the auxiliary variable in the sample and we estimate the linearized variables using the HT weights. The simulation results are reported in Table \ref{comstsi} and \ref{probstsi} and the conclusions are similar to those obtained from the simple
random sampling design without replacement when the size of the sample is $n=200$ which corresponds to the sample sizes inside each stratum.
It is beneficial to use the available auxiliary information when estimating the mean but there is no need to use nonparametric estimators because
they are not more efficient than the GREG estimator. However, for complex parameters, using a GREG estimator to take auxiliary information
into account is not worthwhile in terms of variance while important gains can be made by using B-spline estimators.
The empirical coverage probabilities are all very good except for the Gini index B-spline estimator
of order three with values equal to 89-90\% which confirms the problem of variance underestimation for moderate sample sizes and splines of order three. \\
Based on this example we do not recommend using high order values for  B-spline regression, especially when the
sample sizes are smaller than 500. However, choosing $m=2$ instead of $m=1$ (which corresponds to poststratification) leads to a clear improvement
in terms of efficiency for complex parameters such as the Gini index or the low-income proportion, and we recommend this choice.
\begin{center}
\begin{table}
\caption{\label{comstsi} RRMSE (RB) of HT, GREG and POST, BS(2)
and BS(3) with  $K=2$ - 4}
\centering
\fbox{
\begin{tabular}{*{6}{c}}
\em Parameter & \em  HT&\em GREG& \em POST & \em BS(2)  & \em BS(3)\\
\hline
Mean &   100 (0) & 40 (0) &  73 (0) - 66 (0)  &  40 (0) - 40 (0)  & 40 (0)  - 40 (0)\\
\hline
Gini index &   100 (0) & 93 (0) & 94 (0)   - 88 (0) & 50 (0) - 50 (0) & 55 (1) - 52 (1) \\
\hline
Poverty rate  &  100 (0) & 93 (0) & 88 (0) - 77 (0) & 65 (0) - 64 (0) & 68 (0) - 62 (0)  \\
\hline
\end{tabular}}
\end{table}
\end{center}
\begin{center}
\begin{table}
\caption{\label{probstsi} Coverage probabilities  for HT, GREG and POST, BS(2)
and BS(3) with  $K=2$ - 4}
\centering
\fbox{
\begin{tabular}{*{7}{c}}
\em Parameter & \em  HT&\em GREG& \em POST & \em BS(2)  & \em BS(3)\\
\hline
Mean &   95 & 95 & 95  - 94& 94 - 94 & 93 - 92 \\
\hline
Gini index&  95 & 95 & 95  - 95 & 93  - 93& 89 - 90 \\
\hline
Poverty rate &  94 & 95 & 95  - 95& 95 - 95 &  96  - 95 \\
\hline
\end{tabular}}
\end{table}
\end{center}
\vspace{-15mm}
\section{\baselineskip=10pt Discussion} \label{sec:persp}
In this paper we considered the important problem of nonlinear parameter estimation  in a finite population framework by  taking into account the survey design and a unique auxiliary variable known for all the population units.
Examples of nonlinear parameters are concentration and inequality measures, such as the Gini index or the low-income proportion.
We proposed a general class of substitution estimators that allows us to take into account the auxiliary information via a nonparametric model-assisted approach. The asymptotic variance of this class of estimators was derived, based on broad assumptions, and variance estimators were proposed.
Our main result was that the asymptotic variance depends on the extent to which the auxiliary variable $z_k$ explains the variation in the linearized variable $u_k$.
Because linearized variables of nonlinear parameter are likely to be nonlinearly related to auxiliary information, a nonparametric approach is recommended. The proposed estimators are based on weights that are flexible enough to increase the efficiency  of finite population  totals estimators for any study variable and to allow the consideration of parameters that are more complex than totals.
Moreover, the penalized B-spline estimators were studied in detail, and the theoretical results were confirmed for regression B-spline estimators using one case study.

Our proposal can be extended in several different ways. In particular, further research can extend this proposal to include  multivariate auxiliary information by means of additive models, as in Breidt \textit{et al.} (2005), or single index models as in Wang (2009). \\

\noindent {\bf Acknowledgement:} we are grateful to Patrick Gabriel for his precious help for lemma 1, to Herv\'e Cardot for helpful discussions and to Didier Gazen for his assistance with the simulations.

\section*{\baselineskip=10pt Appendix: assumptions and  proofs}


\appendix
\textbf{Assumptions on functional $T$ and on sampling design.}
{\small{\begin{itemize}
\item[(A1).] The functional $T$ is homogeneous, in that there exists a real number $\alpha>0, $ dependent on $T$ such that  $T(rM)=r^{\alpha}T(M)$ for any real $r>0. $ We assume also that $\lim_{N\rightarrow\infty}N^{-\alpha}T(M)<\infty.$
\item[(A2).] The functional $T$ is  Fr\'echet differentiable at $M/N$; that is, there exists a  functional $T(M/N; \Delta)$ that is linear in $\Delta$ such that $\displaystyle
\left|T\left(\frac{G}{N}\right)-T\left(\frac{M}{N}\right)-T\left(\frac{M}{N}; \frac{G-M}{N}\right)\right|=o\left(d\left(\frac{G}{N},\frac{M}{N}\right)\right)$
 with  $d\left(\frac{G}{N},\frac{M}{N}\right)\longrightarrow 0.$
 \end{itemize}
 We note that the strong assumption of Fr\'echet differentiability can be weakened to  compact or Hadamard  differentiability.  However, for Hadamard  differentiability, functionals are considered with respect to the empirical distribution function and the distance $d_{\mbox{tv}}$ should be replaced by the sup norm.  Supplementary assumptions need to be supposed in order to have the consistency of the estimator of the empirical distribution function. Motoyama and Takahashi (2008) study the asymptotic behavior of Hadamard statistical functionals but only for simple random sampling without replacement.
 \begin{itemize}
\item[(A3).] $\displaystyle\lim_{N\rightarrow\infty}\frac{n}{N}=\pi\in (0, 1). $
\item[(A4).] \begin{itemize}
\item [(a)] $ \overline{\lim} N^{-1}\sum_{U}y_k^2<\infty$ with $\xi$-probability 1.
\item [(b)] $\mbox{sup}_{k\in U}y_k\leq C$ with $C$ a positive constant not depending on $N.$
\end{itemize}
\item[(A5).]
$ \displaystyle \min_{k\in U}\pi_k\geq \lambda$, $\min_{i,k\in U}\pi_{ik}\geq \lambda*$ with  $\lambda, \lambda*$ with some positive constants and $\displaystyle \overline{\lim}_{N\rightarrow\infty} n \ \max_{i\neq k\in U}|\pi_{ik}-\pi_i\pi_k|<\infty. $
\end{itemize}
Assumption (A3) and (A5) deal with first and second order inclusion probabilities and are rather classical in survey sampling theory (see also Robinson and S\"arndal, 1983 and Breidt and Opsomer, 2000).  They are satisfied for many sampling designs. Assumption (A4)-(a) is a regularity condition necessary to get the consistency results. Some results need the stronger assumption (A4)-(b).

\vspace{0.3cm}

\noindent\textbf{Assumptions on B-splines}
\begin{itemize}
\item[(B1).] There exists a distribution function $Q(z)$ with strictly positive density  on $[0,1]$ such that
$\sup_{z\in [0,1]} |Q_N(z)-Q(z)|= o(K^{-1}),$ with $Q_N(z)$ the empirical distribution of $(z_i)_{i=1}^N.$
\item[(B2).]  \begin{itemize}
\item [(a)] $K=o(N)$;
\item [(b)] $K=O(n^a)$ with $a<1/3. $
\end{itemize}
\item[(B3).] $K_p=(K+m-p)(\lambda \tilde c)^{1/(2p)}N^{-1/(2p)}<1$ where $\tilde c=c(1+o(1))$ with $c$ a constant that depends only on $p$ and the design density.
\end{itemize}}
These assumptions are classical in nonparametric regression (Agarwal and Studden, 1980, Burman, 1991, Zhou \textit{et al.}, 1998); (B1) means that asymptotically, there is no sub-interval in $[0,1]$ without  points $z_i$ and (B2) ensures that the dimension of the B-spline basis goes to infinity but not too fast when the population and the sample sizes go to infinity. Assumption (B3) concerns the penalty $\lambda$ as used by Claeskens \textit{et al.} (2009).

\vspace{0.3cm}

\subsection*{Proofs of results from section \ref{sec:nonpaux}}

\vspace{0.3cm}
\noindent\textbf{Proof of Lemma \ref{lemth1}}.
\vspace{0.3cm}
Now, let $h\in{\cal H}$ and let $I_k=1_{\{k\in s\}}$ be the sample membership.  Following the same lines as in Breidt and Opsomer (2000), we have,
\begin{eqnarray*}
E_p\left|\int h\, d\widehat{M}_{HT}/N-\int h\, dM/N\right|^2 & = & N^{-2}\mbox{Var}_p\left(\sum_U\frac{I_k}{\pi_k}h(y_k)\right)\\
& \leq & \left(\frac{1-\lambda}{\lambda N}+\frac{n\mbox{max}_{k\neq l}|\Delta_{kl}|}{\lambda^2n}\right)\frac{1}{N}\sum_Uh^2(y_k)\\
 &\leq & \frac{1-\lambda}{\lambda N}+\frac{n\mbox{max}_{k\neq l}|\Delta_{kl}|}{\lambda^2n}=O(n^{-1})
\end{eqnarray*}
uniformly in $h$ by assumption (A3),(A5) and using the fact that $h\in \cal H. $ \\

\vspace{0.3cm}

\noindent\textbf{Proof of Lemma \ref{lemth2}}. We have
\begin{eqnarray*}
E_p\left|\int h\,d\left(\frac{\widehat M_{np}}{N}\right)-\int h\,d\left(\frac{M}{N}\right)\right|
\leq   E_p\left|\frac{1}{N}\sum_U\left(\frac{I_k}{\pi_k}-1\right)(x_k-\tilde f_{x,k})\right|+E_p\left|\frac{1}{N}\sum_U\left(\frac{I_k}{\pi_k}-1\right)(\tilde f_{x,k}-\hat f_{x,k})\right|
\end{eqnarray*}
From the proof of lemma \ref{lemth1}, we see that the first term from the right-side is of order $O(n^{-1/2})$  uniformly in $h$ because $(1/N)\sum_U(x_k-\tilde f_{x,k})^2\leq (2/N)\sum_U(x^2_k+\tilde f^2_{x,k})\leq 2(1+C)$ by construction of $x_k$ and assumption ($A^*$). The result follows because $|\int h\,d\hat M_{np}/N-\int h\,dM/N|=O_p(n^{-1/2})$ uniformly in $h\in \mathcal H.$\\

\vspace{0.3cm}
\noindent\textbf{Proof of Theorem \ref{nonlinaux}}
Under assumption (A2), we provide a first-order von-Mises (1947) expansion  of $T$ in $\widehat M_{np}/N$ around $M/N,$
\begin{eqnarray*}
T\left(\frac{\widehat M_{np}}{N}\right)=T\left(\frac{M}{N}\right)+\int I T\left(\frac{M}{N},\xi\right)d\left(\frac{\widehat M_{np}}{N}-\frac{M}{N}\right)(\xi)+o\left(d_{\mbox{tv}}\left(\frac{\widehat{M}_{np}}{N},\frac{M}{N} \right)\right).\label{vonmises1}
\end{eqnarray*}
Using the fact that  for a functional of degree $\alpha$ (assumption A1), we have $I T\left(\frac{M}{N},\xi\right)=N^{1-\alpha}\cdot I T\left(M,\xi\right)$ (Deville, 1999), we write
\begin{eqnarray}
N^{-\alpha} T(\widehat M_{np})=N^{-\alpha} T(M)+N^{-\alpha}\int I T\left(M,\xi\right)d(\widehat M_{np}-M)(\xi)+o_p(n^{-1/2})\label{vonmises2}
\end{eqnarray}
since $d_{\mbox{tv}}\left(\widehat{M}_{np}/N,M/N \right)=O_p(n^{-1/2}). $ Now, $u_k=IT(M,y_k)$ and hence, relation (\ref{vonmises2}) becomes,
\begin{eqnarray*}
N^{-\alpha}\left(\widehat\Phi_{np}-\Phi\right) & = &  N^{-\alpha}\left(\sum_sw_{ks}u_k-\sum_Uu_k\right)+ o_p(n^{-1/2}).
\end{eqnarray*}

\subsection*{Proofs of results from section \ref{sec:bsplines}}
\vspace{0.3cm}

We state below several lemmas useful for the proofs of our main results. For a matrix $\mathbf{A}=(a_{ij})_{i,j=1}^p,$  we consider the norm defined by $||\mathbf{A}||_{\infty}=\mbox{max}_{i=1}^p\sum_{j=1}^p|a_{ij}|$ and the trace norm $||\mathbf{A}||_2=(\mbox{trace}(\mathbf{A'}\mathbf{A}))^{1/2}.$ \\

\noindent We denote by $\mathbf{H}_{\lambda}=\frac{1}{N}\mathbf{B}'_U\mathbf{B}_U+\frac{\lambda}{N} \mathbf{D}_p$ and by $\widehat{\mathbf{H}}_{\lambda}=\frac{1}{N}\mathbf{B}'_s\mathbf{\Pi}_s^{-1}\mathbf{B}_s+\frac{\lambda}{N} \mathbf{D}_p$ its estimator.

\begin{lem}\label{lemma_gen} Assume assumptions (B1), (B2)-(a) and (B3). Then,
\begin{enumerate}
\item $||\frac{1}{N}(\mathbf{B}'_U\mathbf{B}_U)||_{\infty}=O(K^{-1}),$ (lemma 6.3 from Agarwal and Studden, 1980). \\
We also have $||(\frac{1}{N}\mathbf{B}'_U\mathbf{B}_U)^{-1}||_{\infty}=O(K),$ (lemma 6.3 from Zhou \textit{et al.}, 1998).\\

\item $||(\frac{1}{N}\mathbf{B}'_U\mathbf{B}_U+\frac{\lambda}{N} \mathbf{D}_p)^{-1}||_{\infty}=||\mathbf{H}^{-1}_{\lambda}||_{\infty}=O(K)$ (lemma 1 from Claeskens \textit{et al.}, 2009)
\end{enumerate}
\end{lem}
\begin{lem}\label{lemma_goga}(Goga, 2005)  Assume (A3), (A4)-(a), (A5) and (B1), (B2)-(a). Then,
\begin{enumerate}
\item $E_p||\frac{1}{N}\left(\mathbf{B}'_s\mathbf{\Pi}_s^{-1}\mathbf{B}_s-\mathbf{B}'_U\mathbf{B}_U\right)||_2^2=O(\frac{1}{n})$\\

\item $E_p||\frac{1}{N}\left(\mathbf{B}'_s\mathbf{\Pi}_s^{-1}\mathbf{y}_s-\mathbf{B}'_U\mathbf{y}_U\right)||^2=O(\frac{1}{n})$
\end{enumerate}
\end{lem}

\vspace{2mm}

\noindent\textbf{Proof of lemma \ref{lemme_thetay}}. When $y_k$ is uniformly bounded (assumption A4,b), we have, using lemma 3 (a) (Goga, 2005) that
\begin{eqnarray}
||\frac{1}{N}\mathbf{B}'_U\mathbf{y}_U||^2\leq \frac{C^2}{N}||\sum_U\mathbf{b}(z_k)||^2\leq \frac{1}{K}\label{borne_BU_YU}
\end{eqnarray}
since for $k,l\in U$ with  $|k-l|>m$ we have $B_j(x_k)B_j(x_l)=0.$

\noindent For (a),  $\tilde{\boldsymbol{\theta}}_{y,\lambda}$ is bounded following Goga (2005),
\begin{eqnarray}
||\tilde{\boldsymbol{\theta}}_{y,\lambda}|| & \leq & ||\mathbf{H}^{-1}_{\lambda}||_{\infty}\cdot||(1/N)\mathbf{B}'_U\mathbf{y}_U||\nonumber\\
 &= & O(K^{-1/2})\label{borne_tilde_theta}
\end{eqnarray}
by lemma \ref{lemma_gen}-(b) and relation (\ref{borne_BU_YU}). Furthermore, we have
\begin{eqnarray}
 & & ||\hat{\boldsymbol{\theta}}_{y,\lambda}-\tilde{\boldsymbol{\theta}}_{y,\lambda}||^2 \nonumber\\
& \leq & ||\widehat{\mathbf{H}}^{-1}_{\lambda}-\mathbf{H}^{-1}_{\lambda}||_{\infty}^2\cdot||\frac{1}{N}\mathbf{B}'_s\mathbf{\Pi}_s^{-1}\mathbf{y}_s||^2+||\mathbf{H}^{-1}_{\lambda}||_{\infty}^2\cdot||\frac{1}{N}\left(\mathbf{B}'_s\mathbf{\Pi}_s^{-1}\mathbf{y}_s-\mathbf{B}'_U\mathbf{y}_U\right)||^2\label{theta_hat_tilde}
\end{eqnarray}
Under the assumption (A4)-(b),  $||\frac{1}{N}\mathbf{B}'_s\mathbf{\Pi}_s^{-1}\mathbf{y}_s||^2$ is bounded by $||\frac{1}{N}\sum_U\mathbf{b}(z_k)||^2=O(K^{-1}).$
We have that
\begin{eqnarray*}
 & &\widehat{\mathbf{H}}^{-1}_{\lambda} -  \mathbf{H}^{-1}_{\lambda} \\
&= & -\mathbf{H}^{-1}_{\lambda}\left(\frac{1}{N}\left(\mathbf{B}'_s\mathbf{\Pi}_s^{-1}\mathbf{B}_s-\mathbf{B}'_U\mathbf{B}_U\right)\right)\left(\mathbf{H}^{-1}_{\lambda}\frac{1}{N}\left(\mathbf{B}'_s\mathbf{\Pi}_s^{-1}\mathbf{B}_s-\mathbf{B}'_U\mathbf{B}_U\right)+\mathbf{I}_{q}\right)^{-1}\mathbf{H}^{-1}_{\lambda}
\end{eqnarray*}
and $||\mathbf{H}^{-1}_{\lambda}\frac{1}{N}\left(\mathbf{B}'_s\mathbf{\Pi}_s^{-1}\mathbf{B}_s-\mathbf{B}'_U\mathbf{B}_U\right)||_{\infty}=o_p(1)$ for $K=O(n^{a})$ with $a<1/3,$ implying that \\
$|| \left(\mathbf{H}^{-1}_{\lambda}\frac{1}{N}\left(\mathbf{B}'_s\mathbf{\Pi}_s^{-1}\mathbf{B}_s-\mathbf{B}'_U\mathbf{B}_U\right)+\mathbf{I}_{q}\right)^{-1}||_{\infty}\leq 1.$ Using lemma \ref{lemma_goga}-(a), we obtain that
\begin{eqnarray*}
E_p||\widehat{\mathbf{H}}^{-1}_{\lambda}-\mathbf{H}^{-1}_{\lambda}||_{\infty}^2& =O\left(\frac{K^4}{n}\right) &\\
\end{eqnarray*}
From lemmas \ref{lemma_gen} and \ref{lemma_goga}, we obtain that
\begin{eqnarray*}
E_p\left(||\mathbf{H}^{-1}_{\lambda}||_{\infty}^2\cdot||\frac{1}{N}\left(\mathbf{B}'_s\mathbf{\Pi}_s^{-1}\mathbf{y}_s-\mathbf{B}'_U\mathbf{y}_U\right)||^2\right)=O\left(\frac{K^2}{n}\right).
\end{eqnarray*}
Finally, we have that
$$
E_p||\hat{\boldsymbol{\theta}}_{y,\lambda}-\tilde{\boldsymbol{\theta}}_{y,\lambda}||^2=O\left(\frac{K^3}{n}\right).
$$

\vspace{0.3cm}

\noindent\textbf{Proof of proposition \ref{lemma_hattBS}}. Consider first $(b). $ Using the same lines as in the proof of lemma \ref{lemth1}  and the fact that $||\mathbf{b}(z_k)||\leq 1$ for all $k\in U$ (Burman, 1991), we obtain that
\begin{eqnarray}
E_p\left|\left|\frac{1}{N}\sum_U\left(\frac{I_k}{\pi_k}-1\right)\mathbf{b}'(z_k)\right|\right|=O(n^{-1/2}).\label{erreur_ht_spline}
\end{eqnarray}
Furthermore,
\begin{eqnarray*}
E_p\left|\frac{1}{N}\left(\hat t_{y,BS}-t^*_{y,diff}\right)\right| &\leq &E_p\left(\left|\left|\frac{1}{N}\sum_U\left(\frac{I_k}{\pi_k}-1\right)\mathbf{b}'(z_k)\right|\right|\cdot||\hat{\boldsymbol{\theta}}_{y,\lambda}-\tilde{\boldsymbol{\theta}}_{y,\lambda}||\right)\\
& \leq & \sqrt{E_p\left|\left|\frac{1}{N}\sum_U\left(\frac{I_k}{\pi_k}-1\right)\mathbf{b}'(z_k)\right|\right|^2\cdot E_p||\hat{\boldsymbol{\theta}}_{y,\lambda}-\tilde{\boldsymbol{\theta}}_{y,\lambda}||^2}\\
& = & O\left(\frac{K^{3/2}}{n}\right)=O\left(\frac{1}{\sqrt{n}}\right)\cdot O((K^3/n)^{1/2})=O\left(\frac{1}{\sqrt{n}}\right)\cdot O((n^{3a-1})^{1/2})\\
& = & o(n^{-1/2})
\end{eqnarray*}
by  (\ref{erreur_ht_spline}) and lemma \ref{lemme_thetay}-(b). Then, the result follows by using the Markov inequality. \\
$(a)$ Now, we consider the error $\hat t_{y,BS}-t_y.$ We write
\begin{eqnarray*}
\frac{1}{N}\left(\hat t_{y,BS}-t_y\right) & = & \frac{1}{N}\left(\hat t_{y,HT}-t_y\right) \\
& - & \frac{1}{N}\sum_U\mathbf{b}'(z_k)\left(\frac{I_k}{\pi_k}-1\right)(\hat{\mathbf{\theta}}_{y,\lambda}-\tilde{\boldsymbol{\theta}}_{y,\lambda})-\frac{1}{N}\sum_U\mathbf{b}'(z_k)\left(\frac{I_k}{\pi_k}-1\right)\tilde{\boldsymbol{\theta}}_{y,\lambda}
\end{eqnarray*}
By assumptions (A3), (A4-b) and (A5), we have that $E_p\left|\frac{1}{N}\left(\hat t_{y,HT}-t_y\right)\right|=O(n^{-1/2}). $ Moreover, using relation (\ref{erreur_ht_spline}) and lemma \ref{lemme_thetay}, (a) we have
%
$E_p\left|\frac{1}{N}\sum_U\left(\frac{I_k}{\pi_k}-1\right)\mathbf{b}'(z_k)\tilde{\boldsymbol{\theta}}_{y,\lambda}\right| =O((K/n)^{1/2})$
which implies that
$$
E_p\left|\frac{1}{N}\left(\hat t_{y,BS}-t_y\right)\right|\leq O(n^{-1/2})+O(K^{3/2}n^{-1})+O(K^{1/2}n^{-1/2})=O((K/n)^{1/2})
$$
by the fact that $(K/n)^{1/2}>n^{-1/2}>K^{3/2}n^{-1}$ using assumption (B2).

\vspace{0.3cm}

\noindent\textbf{Proof of Theorem \ref{thm:bspline}}.
 (a) We check that assumptions ($A^*$) and ($A^{**}$) are fulfilled. We have $\tilde{\boldsymbol{\theta}}_{x,\lambda}=\mathbf{H}^{-1}_{\lambda}(\sum_U\mathbf{b}(z_k)x_k/N)$ with $|x_k|=|h(y_k)|\leq 1$ for all $k\in U.$ Following (\ref{borne_BU_YU}) and (\ref{borne_tilde_theta}), we obtain that $||\tilde{\boldsymbol{\theta}}_{x,\lambda}||=O(K^{1/2})$ uniformly in $h$ and 
 \begin{eqnarray}
\frac{1}{N}\sum_U\tilde f^2_{x,k}=\frac{1}{N}\tilde{\boldsymbol{\theta}}'_{x,\lambda}\mathbf{B}'_U\mathbf{B}_U\tilde{\boldsymbol{\theta}}_{x,\lambda}\leq ||\tilde{\boldsymbol{\theta}}_{x,\lambda}||^2||\frac{1}{N}\mathbf{B}'_U\mathbf{B}_U||_{\infty}=O(1),
\end{eqnarray}
uniformly in $h. $
Now, we check the assumption ($A^{**}$), namely
$E_p\left|\frac{1}{N}\sum_U\left(\frac{I_k}{\pi_k}-1\right)(\tilde f_{x,k}-\hat f_{x,k})\right|=O(n^{-1/2})$ uniformly in $h.$

We have
\begin{eqnarray}
E_p\left|\frac{1}{N}\sum_U\left(\frac{I_k}{\pi_k}-1\right)(\tilde f_{x,k}-\hat f_{x,k})\right|&\leq & E_p\left(\left|\left|\frac{1}{N}\sum_U\left(\frac{I_k}{\pi_k}-1\right)\mathbf{b}'(z_k)\right|\right|\cdot ||\tilde{\boldsymbol{\theta}}_{x,\lambda}-\hat{\boldsymbol{\theta}}_{x,\lambda}||\right)\nonumber\\
 &\leq & \sqrt{E_p\left|\left|\frac{1}{N}\sum_U\left(\frac{I_k}{\pi_k}-1\right)\mathbf{b}'(z_k)\right|\right|^2\cdot E_p||\tilde{\boldsymbol{\theta}}_{x,\lambda}-\hat{\boldsymbol{\theta}}_{x,\lambda}||^2}\nonumber
\end{eqnarray}
The first term from the right-side does not depend on  $h$ and is of order $O(n^{-1})$ (equation (\ref{erreur_ht_spline})).  For the second term from the right-side, we can use the proof of lemma (\ref{lemme_thetay}), more exactly the equation (\ref{theta_hat_tilde}), and the fact that $\mbox{sup}_{k\in U}|h(y_k)|\leq 1$ to obtain
$$
E_p||\tilde{\boldsymbol{\theta}}_{x,\lambda}-\hat{\boldsymbol{\theta}}_{x,\lambda}||^2=O\left(\frac{K^3}{n}\right)\quad \mbox{uniformly in h}.
$$
Finally, we obtain  that $E_p\left|\frac{1}{N}\sum_U\left(\frac{I_k}{\pi_k}-1\right)(\tilde f_{x,k}-\hat f_{x,k})\right|=o(n^{-1/2})$ for $K=O(n^a)$ with $a<1/3.$

\vspace{0.3cm}

\noindent (b) We write equation (\ref{approxbspline}) as follows:
\begin{eqnarray*}
N^{-\alpha} (t^*_{u,BS}-t^*_{u,\mbox{\tiny diff}})= N^{-\alpha}\sum_U \left(\frac{I_k}{\pi_k}-1\right) \mathbf{b'}(z_k)(\hat {\boldsymbol\theta}_{u,\lambda}-\tilde {\boldsymbol\theta}_{u,\lambda})=o_p(n^{-1/2})
\end{eqnarray*}
 because $N^{-1}\sum_U \left(\displaystyle\frac{I_k}{\pi_k}-1\right) \mathbf{b'}(z_k)=O_p(n^{-1/2})$  (equation (\ref{erreur_ht_spline})) and $N^{-\alpha+1}(\hat
{\boldsymbol\theta}_u-\tilde {\boldsymbol\theta}_u)=O_p(K^{3/2}n^{-1/2})$ by lemma \ref{lemme_thetay}.
\vspace{0.3cm}

\noindent\textbf{Proof of Theorem \ref{estim_var_nonparam}}.  The proof follows the same basic steps as in Theorem 3 from Goga \textit{et al.} (2009) and result 4 from Chaouch and Goga (2010). Let
$$A_N=\widehat {V}_p(\widehat{\Phi}_{np})-\widehat{AV}_{p}(\widehat{\Phi}_{np}), \quad B_N=\widehat{AV}_{p}(\widehat{\Phi}_{np})-AV_p(\widehat{\Phi}_{np})
$$
with $\widehat{AV}_{p}(\widehat{\Phi}_{np})$ given by (\ref{varht})  and let also $c_{kl}=\displaystyle \frac{\Delta_{kl}}{\pi_{kl}}\frac{I_k}{\pi_k}\frac{I_l}{\pi_l}$. Furthermore, the quantity $A_N$ can be written as
 \begin{eqnarray*}
 A_N & =  & \sum_U\sum_U c_{kl}(\hat e_{ks}\hat e_{ls}-e_{ks}e_{ls})\\
  & = &  \sum_U\sum_U c_{kl}(\hat e_{ks}-e_{ks})(\hat e_{ls}-e_{ls})+2\sum_U\sum_U c_{kl}(\hat e_{ks}-e_{ks})e_{ls}\\
  & = & A_{1N}+A_{2N}
 \end{eqnarray*}
\noindent Now,
\begin{eqnarray*}
\frac{n}{N^{2\alpha}}|A_{1N}|\leq \frac{1-\lambda}{\lambda^2}\frac{n}{N^{2\alpha}}\sum_U(\hat e_{ks}-e_{ks})^2+\frac{n\max |\Delta_{kl}|}{\lambda^2\lambda^*N^{2\alpha-1}}\sum_U(\hat e_{ks}-e_{ks})^2=o_p(1)
\end{eqnarray*}
by assumptions  (A3) and (A5). Using the same arguments as above, we obtain $nN^{-2\alpha}|A_{2N}|=o_p(n^{-1}).$ Hence, $nN^{-2\alpha}|A_{N}|=o_p(n^{-1})$  and the result then follows because $nN^{-2\alpha}B_N=o_p(1)$
$$
\left|\frac{n}{N^{2\alpha}}(\widehat {V}_p(\widehat{\Phi}_{np})-AV_p(\widehat{\Phi}_{np}))\right|\leq \frac{n}{N^{2\alpha}}(|A_N|+|B_N|).
$$

}


\end{document}